\documentclass[twocolumn,prb,groupedaddress,showpacs,superscriptaddress]{revtex4-2}
\usepackage[utf8]{inputenc}
\usepackage{slashed}
\usepackage{hyperref}
\usepackage{hyperref}
\usepackage{booktabs}
\usepackage{upgreek} 

\usepackage{float}
\usepackage{graphicx}
\usepackage{epstopdf}
\usepackage{tikz-cd}
\usepackage{quiver}
\usepackage{svg}

\usepackage{amscd,amssymb,amsmath,latexsym,bm}
\usepackage{amsmath,amsthm,amssymb}
\usepackage{amsfonts}
\usepackage[mathcal,mathscr]{euscript}
\usepackage{braket}

\newtheorem{theorem}{Theorem}

\newtheorem{proposition}[theorem]{Proposition}
\newtheorem{remark}[theorem]{Remark}
\newtheorem{definition}[theorem]{Definition}

\newcommand{\CM}{{\mathbb C}}
\newcommand{\NM}{{\mathbb N}}

\newcommand{\RM}{{\mathbb R}}
\newcommand{\SM}{{\mathbb S}}
\newcommand{\TM}{{\mathbb T}}
\newcommand{\ZM}{{\mathbb Z}}

\newcommand{\C}{\mathbb{C}} 

\newcommand{\Z}{\mathbb{Z}}
\newcommand{\T}{\mathbb{T}}
\newcommand{\Aa}{{\mathcal A}}

\newcommand{\Bb}{{\mathcal B}}

\newcommand{\Gg}{{\mathcal G}}

\newcommand{\Cc}{{\mathcal C}}

\newcommand{\Ll}{{\mathcal L}}

\newcommand{\Hh}{{\mathcal H}}

\newcommand{\id}{\mathrm{id}}

\begin{document}

\title{Intrinsic Chiral HOTI Stabilized by a Single Diagonal Reflection}

\author{Danilo Polo Ojito}
\email{danilopolo@javeriana.edu.co}
\address{Department of Mathematics, Pontificia Universidad Javeriana, Bogota, Colombia
}

\author{Jackson Saunders}
\email{Jsaunders@u.nus.edu}
\address{Department of Physics and Engineering Physics, Fordham University, 441 East Fordham Rd., Bronx, NY 10458,
USA}
\address{Department of Physics, National University of Singapore, 2 Science Drive 3, Singapore 117551}

\author{Ellis Bender}
\email{ejb1@fordham.edu}
\address{Department of Physics and Engineering Physics, Fordham University, 441 East Fordham Rd., Bronx, NY 10458,
USA}

\author{Camelia Prodan}
\email{cprodan@fordham.edu}
\address{Department of Physics and Engineering Physics, Fordham University, 441 East Fordham Rd., Bronx, NY 10458,
USA}

\author{Emil Prodan}
\email{prodan@yu.edu}
\address{Department of Physics and Department of Mathematical Sciences, Yeshiva University, New York, NY 10016, USA
}
\date{\today}

\begin{abstract} Higher-order topological insulators and superconductors, under combined point and fundamental symmetries, display topological modes supported by boundaries of higher codimensions, such as a corner mode. In this work, we explain a theoretical device that enables us to claim that there exists {\it one and only one} intrinsic chiral-symmetric higher-order topological phase for the quarter-plane geometry with point-symmetry given by reflection about the diagonal of the corner. Furthermore, we supply a minimal tight-binding model of this phase and design an acoustic crystal whose dynamics reproduce that of the model. The robust properties of the corner mode are investigated by COMSOL simulations.
\end{abstract}

\maketitle

\section{Introduction}
 In their seminal works, Benalcazar, Bernevig, and Hughes \cite{BenalcazarScience2017, BenalcazarPRB2017}  discovered a new class of crystalline topological phases exhibiting electronic dynamics confined not only along edges or surfaces, but also on ``boundaries of boundaries," such as hinges and corners. In particular, they demonstrated that certain point-groups can stabilize quantized bulk multipole moments, inducing topologically protected modes localized at intersections of boundaries. Their predictions were confirmed shortly afterward using laboratory models based on microwave circuits \cite{PetersonNat2018} and pononic crystals \cite{SerraGarciaNature2018}. These ideas were subsequently expanded by Schindler {\it et al.} in \cite{SchindlerSciAdv2018}, giving rise to the notion of higher-order topological insulators (HOTIs). Under space terminations that preserve a subgroup of the crystal’s space-group symmetries, HOTIs exhibit topological modes localized on boundary features of co-dimension $n>1$ relative to the bulk. For example, a flat boundary has co-dimension 1, a hinge has co-dimension 2 and a corner has co-dimension 3. Ever since, HOTI has been a very active research topic, explored theoretically and experimentally for a wide variety of crystalline symmetries with or without other fundamental symmetries  \cite{XieNatRevPhys2021}, and even beyond the crystalline setting \cite{YangJPCM2024}.
 


From a mathematical-physics point of view, HOTIs presented several outstanding challenges: (1) The bulk topology and the predicted boundary phenomena, strictly speaking, exist only in the thermodynamic limit, for otherwise the expected boundary modes couple through their evanescent tails and the spectra become gapped. In that infinite-size limit, how are we going to remember (aka encode) how the crystal was terminated? (2) The notion of gappable edges or surfaces, needed to expose the topological modes supported by their higher co-dimensional defects, make good sense at the intuitive level. It infers that there are no topological obstructions against deforming a bulk model and the boundary conditions until the dynamics near a specific co-boundary becomes spectrally gapped. Can this notion be formulated within the established language of topology, which first and foremost calls for a clear identification of the relevant topological spaces where those deformations occur? (3) HOTI phenomena have an intrinsic character, {\it i.e.} they are protected by bulk's topology and symmetries: Different compatible boundary conditions can shuffle, add or remove boundary modes but never fully destroy them. (For equally interesting extrinsic HOTIs, see \cite{HayashiLMP2021}.) Can this intrinsic character be formulated within the established language of topology?

Operator algebras, and in particular $C^\ast$-algebras, have a long and fruitful history of interactions with condensed matter physics. In the following, we summarize a sequence of such interactions that led to affirmative answers for the three stated challenges. First and foremost, $C^\ast$-algebras come equipped with natural topologies that can be used naturally to encode locality and continuity of the physical Hamiltonians under various deformations, supplying in the same time a framework to perform computations in real-space and directly in the thermodynamic limit \cite{Bellissard1,Bellissard2,Bellissard3,ProdanAMRX2013}. Their $K$-theories are devices that sort gapped Hamiltonians in stable homotopy equivalence classes \cite{Bellissard4,KellendonkRMP1995}. Furthermore, the bulk-boundary correspondence principle for flat boundaries of class A insulators has found an elegant formulation in the operator algebraic setting. Indeed, it was shown in \cite{KellendonkRMP2002} that the half-space Hamiltonians, with all physically sound boundary conditions, form a $C^\ast$-algebra. Inside this algebra, one can naturally identify an ideal containing all potentials responsible for various boundary conditions, which enters the exact sequence
\[
\mbox{boundary alg.} \hookrightarrow \mbox{half-space alg.}  \twoheadrightarrow \mbox{bulk alg.} 
\]
In turn, this sets in motion standard exact sequences between corresponding $K$-theories, creating precise bridges between the topologies of the bulk and of the boundary. In particular, one can identify {\it all} bulk models which generate topological boundary states when restricted to half space, regardless of the boundary condition. See \cite{BourneAHP2017} for an extension of these ideas to the full classification table of topological insulators and superconductors.

Besides flat boundaries, there are many other elementary geometric defects in condensed matter systems \cite{MerminRMP1979}, which can support topological states too \cite{TeoPRB2010,TeoARCMP2017}. While \cite{TeoPRB2010} developed a framework based on a mixed momentum-real spaces approach, the operator algebraic framework enables a clean real-space formulation \cite{ProdanJPA2021,DET2}, which {\it e.g.} can incorporate disorder. As it was pointed out in \cite{ProdanJPA2021}, the key to a bulk-defect correspondence is the structure of the phason space $\Xi_\Ll$ of a lattice $\Ll$ in the presence of an elementary geometric defect (see Sec.~\ref{Sec:PhasonSpace}). In a nutshell, the phason space contains all the different images of the same lattice as seen by an observer located at different points in space, and it is this phason space that determines the $C^\ast$-algebra of physical Hamiltonians (see Sec.~\ref{Sec:HamAlg}). In the presence of an elementary geometric defect, the phason space decomposes as $\Xi_\Ll = \Xi_\Ll^\infty \cup \Xi_\Ll^c$, where $\Xi_\Ll^\infty$ encodes the images of the lattice seen by an observer asymptotically far from the defect. By a standard mechanism for (groupoid) $C^\ast$-algebras \cite{WilliamsBook2}, this splitting prompts again an exact sequence 
sequence
\[
\mbox{defect alg.} \hookrightarrow \mbox{full alg.}  \twoheadrightarrow \mbox{bulk alg.}
\]
which again trickles-down to the level of $K$-theory, enabling one to identify {\it all} bulk models which generate topological defect states for a given geometry.

Higher-order boundaries, such as hinges and corners are not elementary geometric defects. Indeed, in the presence of such defects, the phason space has a rich structure, specifically, it accepts a unique maximal filtration 
\[
\{\Ll_{\rm bulk}\}= \Xi_0 \subset \Xi_1 \subset \cdots \subset \Xi_d =\Xi_{\Ll}
\]
by closed and invariant subsets that are compatible with the point-symmetry of the crystal (see Sec.~\ref{Sec:DefectAlgebras}). Here, $\Xi_r\setminus \Xi_{r-1}$ contains the images of the lattice as seen by an observer close to a defect of co-dimension $r$ and asymptotically far from all other defects. This sets in motion a long sequence of $C^\ast$-algebras that can be naturally associated with the dynamics near defects with specified co-dimensions. In turn, this induces a spectral sequence at the level of $K$-theories and, one particular derivation in this sequence gives the affirmative answers to all challenges we stated above \cite{OjitoCMP2025} (see Sec.~\ref{Sec:BulkCorner}). 

The outcome of the operator theoretic framework is a device that can identify {\it all} non-trivial higher-order bulk-defect correspondences compatible with given symmetries and boundary geometry. The scope of this paper is to demonstrate how these operator theoretic devices developed in \cite{OjitoCMP2025} can be used as engines for discovery in the HOTI program. At the same time, we aim to translate the formalism into a language familiar to the materials science community, using the simple case of a (intrinsic) HOTI stabilized by chiral symmetry and just one reflection $(x_1,x_2) \mapsto (x_2,x_1)$ along the diagonal intersecting a corner. While we chose this setting for its simplicity, to authors' knowledge, it was actually never explored before. For example, the original works \cite{BenalcazarScience2017, BenalcazarPRB2017} consider two separate reflections and same in \cite{SchindlerNeupert}, \cite{SarkarPRB2023} considers an additional time reversal symmetry, and the chiral HOTIs identified in \cite{BenalcazarPRB2022,WangPRL2023} are extrinsic (boundary-obstructed). 

In the stated setting, we introduce and characterize the phason space in the presence of a corner, as well as the algebras capturing all physically sound Hamiltonians. Using the structure of the phason space and of the mentioned algebras, we formulate the notion of gappable bulk models at the edges, and define an intrinsic bulk-corner correspondence map. Lastly, we evaluate this map and show that the intrinsic chiral topological states for the corner geometry have a $\ZM_2$-classification hence, up to stable equivariant homotopies, there is one and only one elementary HOTI model for this combination of geometry and symmetry. We supply the minimal tight-binding model for it.

In the second part, we use coupled-mode theory for acoustic resonators to demonstrate the emergence of a topological corner mode in an acoustic crystal, which is built and characterized in COMSOL \cite{COMSOL}. Almost as a rule, even the minimal tight-binding models of HOTIs are still challenging for laboratory implementations. As we shall see, our model requires four resonators per cell and eight connections per resonator, of which several connect second-nearest neighbors. The implementation relies on the so-called H-resonators \cite{saunders2025acousticmetamaterialspositivenegative}, already proven advantageous when dealing with complex structures \cite{WuSciB2024}, but the couplings require several engineering innovations. Indeed, the complexity of the minimal model forces on us to work with acoustic bridges of different lengths, some exceeding the wavelength of sound at the probed frequencies. A controlled procedure to engineer such couplings, inspired from \cite{WangPRL2023}, is explained in Sec.~\ref{Sec:Implementation}.

\section{HOTI by Operator Algebras}

Our main task is to convince the reader that what we are presenting here is not a convenient mathematical framework, but that the operator algebras we are going to reveal accurately describe the dynamics of the truncated lattices without simplifying assumptions. We will be dealing with discrete collections of uniformly separated points of $\RM^2$, which are closed subsets of $\RM^2$ where the lattices live. The spaces of closed subsets of topological spaces, sometimes called hyperspaces, are well studied in mathematics \cite{NadlerBook} and prominently featured in image analysis \cite{BarnsleyBook}. Consideration of the topological space $\Cc(\RM^2)$ of closed subsets of $\RM^2$ (endowed with Fell's topology \cite{FellPAMS1962}) enables us to represent a lattice $\Ll$ as a single point of $\Cc(\RM^2)$, hence to give sense to statements like ``continuous deformations of lattice" or ``depends continuously on the lattice." This is the point of departure for the operator theoretic approach.

\subsection{Emergence of the phason space from dynamical considerations}

\begin{figure}[t!]
\center
	\includegraphics[width=0.99\linewidth]{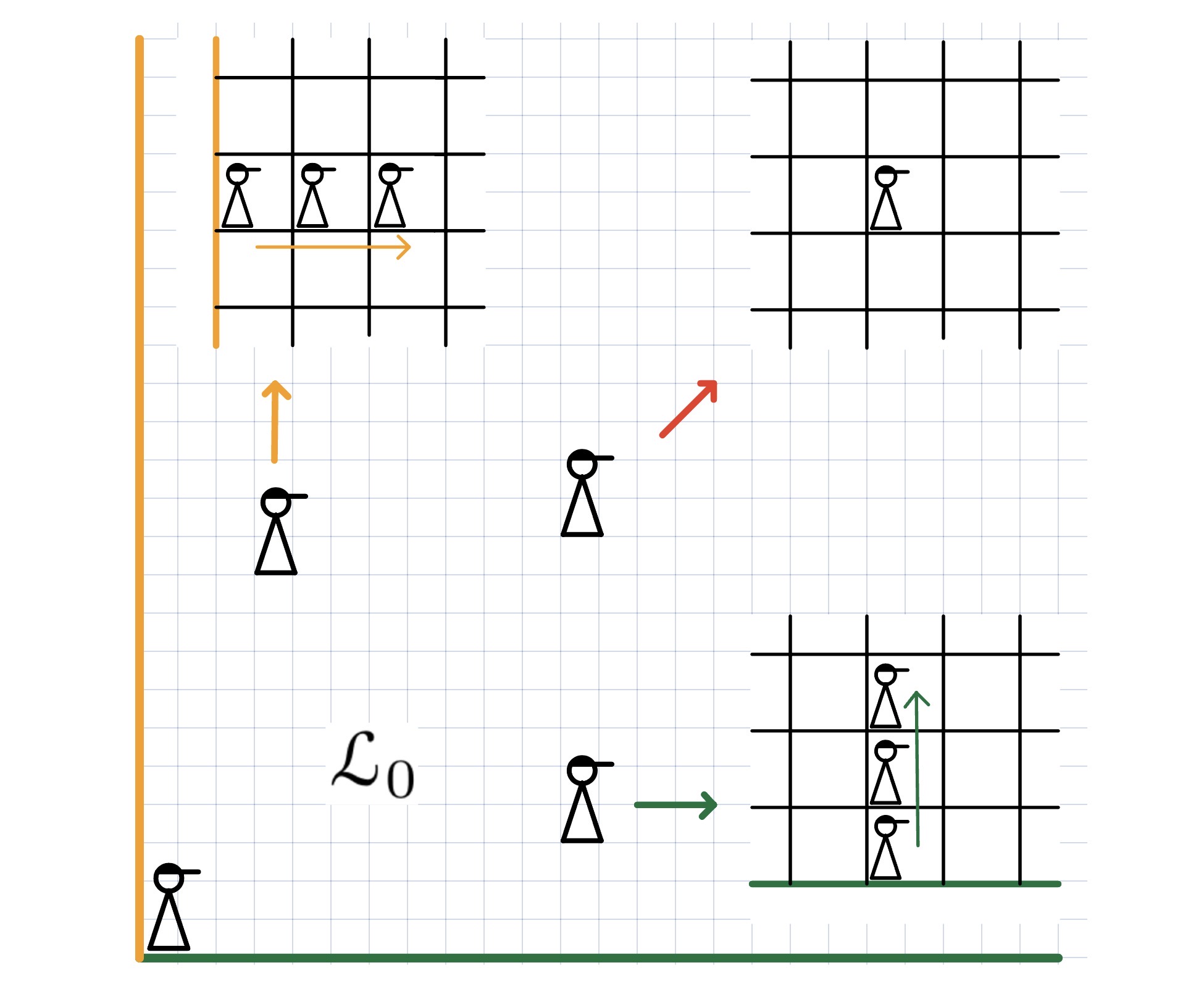}
	\caption{\small An observer inspecting the lattice from different points, with emphasis on the asymptotic regions.}
	\label{Fig:PatternInsp}
\end{figure}

We can describe electronic and acoustic crystals simultaneously if we view the acoustic cavities and their resonant modes as synthetic atoms with synthetic orbitals. For the start, let us consider a generic lattice $\Ll$, which can have boundaries. The Hilbert space of the collective dynamics is spanned by $|\alpha,x\rangle$ representing the orbital $\alpha=1,\ldots,N$ carried by the atom $x \in \Ll$, hence we are dealing with $\CM^N \otimes \ell^2(\Ll)$. Like any bounded linear operator over this Hilbert space, the Hamiltonian generating the collective dynamics takes the tight-binding form 
\[
H_\Ll = \sum_{x,x' \in \Ll} w_{x,x'}(\Ll) \otimes |x\rangle \langle x' |.
\]
Here, $w$'s are $N \times N$ coupling matrices acting on the $\alpha$ part of the mode $|\alpha,x\rangle$ and, by writing $w_{x,x'}(\Ll)$, we indicate that the entries of these matrices are fully determined by the lattice $\Ll$. In other words, we have a well defined continuous map $\Ll \mapsto H_\Ll$ from lattices to Hamiltonians. Furthermore, as a direct consequence of Galilean invariance of the physical laws governing the mode-coupling, we have 
\begin{equation}\label{Eq:Equiv}
w_{x-a,x'-a}(\Ll-a) = w_{x,x'}(\Ll)
\end{equation}
for all $ a\in \RM^d$ and $x,x' \in \Ll$. These relations say that a rigid motion of the lattice will go un-noticed by someone observing the dynamics from a frame tide to the lattice. As a result, the dynamical matrices have a particular structure
\begin{equation}\label{Eq:HL}
    H_\Ll = \sum_{x,x' \in \Ll} w_{0,x'-x}(\Ll-x) \otimes |x\rangle \langle x'|.\vspace{-0.2cm}
\end{equation}
This form automatically encodes the Galilean invariance of the physical laws, which is a separate mater from the symmetries of the lattice itself. 

Now, note that $w$'s in \eqref{Eq:HL} are evaluated on the subset 
\begin{equation}\label{Eq:Transversal}
\{\Ll -x, \ x \in \Ll\} \subset \Cc(\RM^2).
\end{equation}
The closure $\Xi_{\Ll}$ of this subset in the topology of $\Cc(\RM^2)$ coincides with the phason space in all cases studied by materials scientists, and in particular for quasi-crystals \cite{KellendonkAHP2019}. In the mathematics of patterns, $\Xi_{\Ll}$ is called the transversal of $\Ll$ \cite{KellendonkRMP1995} because it is a genuine Poincar\'e section of the orbit $\{\Ll -a, \ a \in \RM^2\}$ inside the pattern space $\Cc(\RM^2)$. 

\subsection{The phason space in the presence of a higher-order boundary}\label{Sec:PhasonSpace}

We now consider $\Ll_0$ to be the quarter-plane lattice shown in Fig.~\ref{Fig:PatternInsp}, as seen by an observer sitting right at its tip. Our task is resolve the topological type of the transversal $\Xi_{\Ll_0}$, {\it i.e.} of the closure of the set \eqref{Eq:Transversal} for this $\Ll_0$, which lives in the space of patterns $\Cc(\RM^2)$. It will be a discrete set with accumulation points. Note that the operation $\Ll_0 -x$ brings resonator $x$ to the origin of $\RM^2$, so we can understand $\Ll_0 - x$ as the lattice seen by an observer located at the point $x$ of the lattice. This small observation is very much the key for mapping the transversals of patterns. 

\begin{figure}[t!]
\center
	\includegraphics[width=0.99\linewidth]{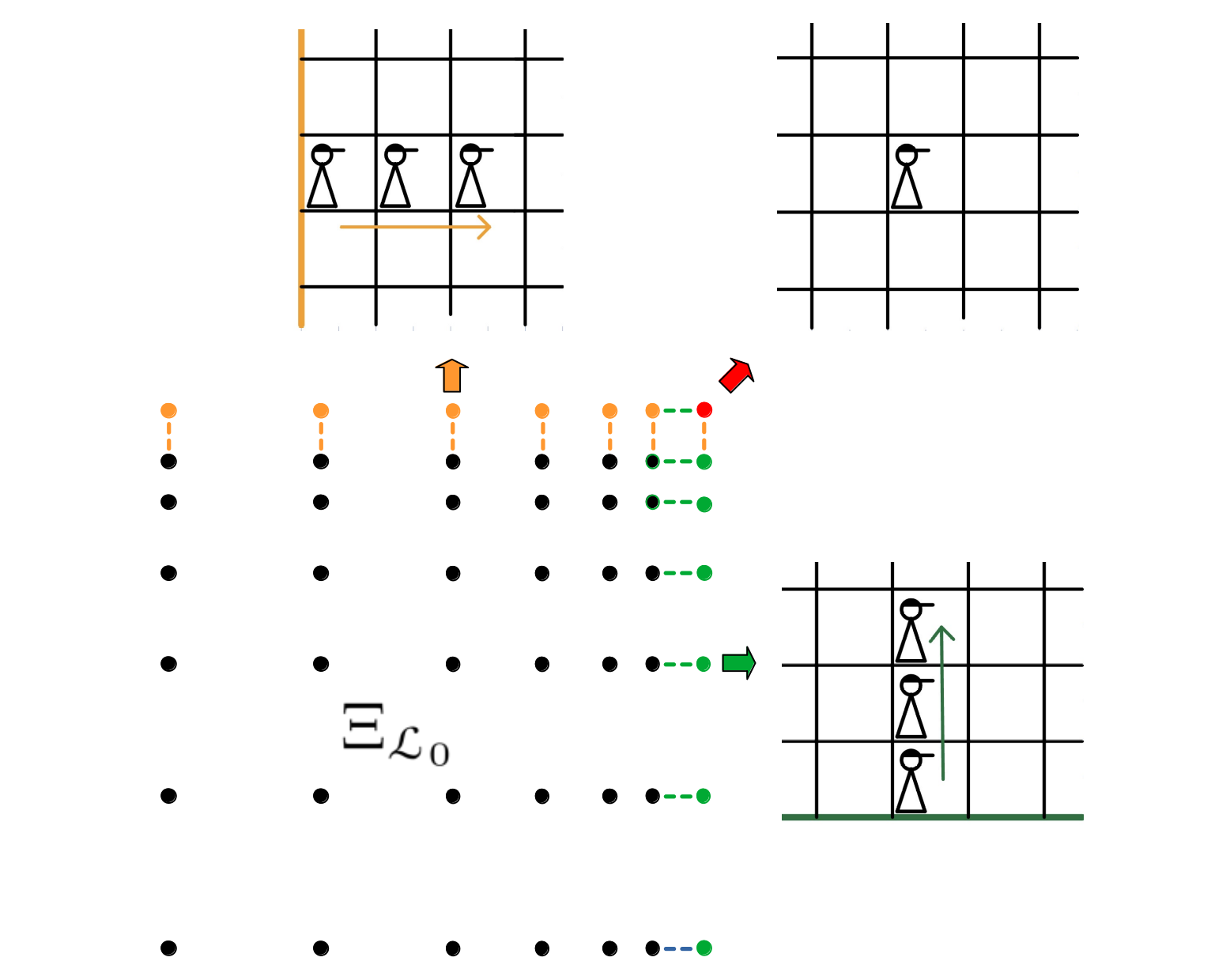}
	\caption{\small Mapping the phason space of $\Ll_0$.}
	\label{Fig:Transversal}
\end{figure}

Now, if the observer steps on the lattice in any direction while the corner is in its field of view, the observer will perceive a different lattice every time. In the space $\Cc(\RM^2)$ of patterns, we put down these (shifted) lattices as the discrete black points seen in Fig.~\ref{Fig:Transversal}. However, if the observer steps in a straight line to the right, the pattern of points will start to look more and more like in inset A of Fig.~\ref{Fig:PatternInsp}. When the corner falls beyond the horizon, the observer reached a limit point, which we represent as a green dot in Fig.~\ref{Fig:Transversal}. There are many green dots because the observer still sees the horizontal edge, hence, while stepping vertically away from the edge, the observer will still perceive different patterns of points. It should be obvious that the observer will become trapped in these limiting patterns, since it will be impossible to walk back and bring the corner in the field of view. 

A similar story will unfold if the observer walks vertically in a straight lines. In this case, the observer reaches the limiting patterns shown in inset B of Fig.~\ref{Fig:PatternInsp}, which are represented by the orange dots in the space of patterns shown in Fig.~\ref{Fig:Transversal}. However, if the observer steps diagonally on the lattice, the corner as well as the edges will fall beyond the horizon and the observer will perceive the pattern shown in the inset C of Fig.~\ref{Fig:PatternInsp}. This is the bulk lattice $\Ll_b$, which we represent as the red dot in the space of patterns. This limiting pattern can be also reached by walking away from the edges, so the transversal must look like as depicted in Fig.~\ref{Fig:Transversal}.

\subsection{The algebra of physical Hamiltonians}\label{Sec:HamAlg}

For a bulk system $\Ll_b - x = \Ll_b$, hence the equivariance relation~\eqref{Eq:Equiv} forces the following structure on the physical Hamiltonians
\[
H_{\Ll_b} = \sum_{x,x' \in \Ll_b} w_{x'-x} |x\rangle \langle x' | = \sum_{q \in \Ll_b} w_{q} \otimes \sum_{x \in \Ll_b} |x\rangle \langle x+q |.
\]
The last sum is nothing but the shift operator $S^q = S_1 ^{q_1}S_2^{q_2}$ on the lattice, hence 
\begin{equation}\label{Eq:HBulk}
H_{\Ll_b} = \sum_{q \in \Ll_b} w_q \otimes S^q = \sum_{q \in \Ll_b} w_q \otimes S_1^{q_1} S_2^{q_2}
\end{equation}
and on the right we see the standards elements of the group algebra $\CM \ZM^2$, as cannonically represented on $\ell^2(\Ll_b)$, tensored by $M_\infty$. The latter is the algebra of infinite-rank matrices that can be approximated by finite rank matrices. Any finite-rank matrix can be naturally embedded in $M_\infty$, and having them inside one space it enables us to turn on and off orbitals, {\it e.g.} during a material deformation, as we should because \eqref{Eq:HL} are only low energy approximations of the exact Hamiltonians.

In the presence of a boundary, the $\ZM^2$-symmetry of the lattice is lost, but the systems still obey the equivariance relation~\eqref{Eq:Equiv}. To take advantage of it, we must pass from groups to the more general notion of groupoids. Specifically, for a fixed aperiodic architecture $\Ll_0$,  its groupoid $\Gg_{\Xi_{\Ll_0}}$ consists of triples 
\begin{equation}\label{Eq:Triples}
\tau=(\Ll-x, x, \Ll),\quad \Ll \in \Xi_{\Ll_0} \ \& \ x \in \Ll,
\end{equation} 
and two such tripples $( \Ll'-x',x',\Ll')$ and $(\Ll-x, x, \Ll)$ are composable if and only if $\Ll' = \Ll-x$, in which case
\[
(\Ll'-x',x',\Ll')(\Ll -x, x, \Ll)=(\Ll-x-x', x+x',\Ll).
\]
There is also a natural inversion,
\[
(\Ll-x, x, \Ll)^{-1}=(\Ll, -x, \Ll-x).
\]
Furthrmore, $\Gg_{\Xi_{\Ll_0}}$ inherits a topology from the space $\Xi_{\Ll_0} \times \RM^2 \times \Xi_{\Ll_0}$ where the triples live. One should be fully aware that the topology and the algebraic structure of the groupoid are equally important.

\begin{remark}{\rm 
For orientation, if we replace $\Ll_0$ by the bulk lattice $\Ll_b$, then $\Ll_b -x =\Ll_b$ and, as such, the phason space contains just one point, $\Ll_b$ itself, and the groupoid reduces to the group $\Ll_b \simeq \ZM^2$.}
\end{remark}

Note that each triple takes an $\Ll \in \Xi_{\Ll_0}$ and sends it into $\Ll -x$, which is also a point of $\Xi_{\Ll_0}$. This is called the groupoid action on the phason space and will play an important role in the following. Indded, we will be seeking closed subsets of $\Xi_{\Ll_0}$ that are invariant under this groupoid action.

Now, like groups, a groupoid accepts a natural algebra. Its elements are complex-valued function with compact support, defined over the groupoid itself. Thus, in our context, we are talking about functions mapping triples $\tau$ to $\CM$, and we will denote this algebra as $\CM \Gg_{\Xi_{\Ll_0}}$, as one will do for group algebras. Addition is just the ordinary addition for complex-valued functions, while the product is given by the convolution
\[
(f \cdot f')(\tau) = \sum_{\tau_1 \tau_2 =\tau} f(\tau_1) f'(\tau_2)
\]
One can verify that this product reduces to the one of the group algebra $\CM \ZM^2$ if we replace $\Ll_0$ with the bulk lattice.

The algebra itself is a linear space, so we can generate a canonical representation of $\CM \Gg_{\Xi_{\Ll_0}}$ by letting it act on itself from the left. This representation, called the left-regular representation, breaks down into a direct sum indexed by the phason space $\pi_L = \bigoplus_{\Ll \in \Xi_{\Ll_0}} \pi_{\Ll}$, were the Hilbert space of $\pi_\Ll$ is spanned by $|x\rangle$ with $x \in \Ll$ and
\begin{equation}\label{Eq:PL}
    \pi_\Ll(f) = \sum_{x,x'\in \Ll} f(\Ll-x+x',x'-x,\Ll)|x\rangle \langle x'|.
\end{equation}
The punch line is that, after we take the tensor with $M_\infty$ and make the choice 
\[
f(\Ll-x,x,\Ll) = w_{0,x}(\Ll)
\]
for all $\Ll \in \Xi_{\Ll_0}$ and $x \in \Ll$, expressions \eqref{Eq:HL} and \eqref{Eq:PL} coincide. Thus, $M_\infty \otimes \CM \Gg_{\Xi_{\Ll_0}}$ can produce with its standard family of representations all generators of the collective dynamics, as observed from any point of the lattice $\Ll_0$. In the same time, we can be sure that any self-adjoint element from this algebra produces a physically sound finite-range dynamics, because we insisted that the functions $f$ have compact supports.

As a last step, let us observe that, even for finite-range Hamiltonians, the Green functions and spectral projectors are not of finite-range type, hence not inside $M_\infty \otimes \CM \Gg_{\Xi_{\Ll_0}}$. To generate the minimal algebra that includes all of that, we close $\CM \Gg_{\Xi_{\Ll_0}}$ inside the algebra of linear operators over the Hilbert space of the left-regular representation, finally leading us to the $C^\ast$-algebra $M_\infty \otimes C^\ast \Gg_{\Xi_{\Ll_0}}$ of physical Hamiltonians.

\subsection{Symmetries}

We can quickly introduce the general framework for mixed space and fundamental symmetries and then particularize to the case of just one space symmetry, the reflection $\sigma_r$ along the first diagonal $(x_1,x_2) \mapsto (x_2,x_1)$, augmented by chiral symmetry. Henceforth, let $\Sigma$ be a finite group of isometry $\sigma$ of $\RM^2$ which leave $\Xi_{\Ll_0}$ invariant. Any such $\sigma$ induces an automorphism of algebra $C^\ast \Gg_{\Ll_0}$,
\[
[\alpha_{\sigma}(f)](\Ll - x,x,\Ll) = f(\sigma\cdot (\Ll -x),\sigma \cdot x,\sigma \cdot \Ll).
\]
The fundamental symmetries are generated by time-reversal $T$ and particle-hole exhange $P$ and, at the first step, they are accounted for by the larger group $\bar \Sigma = \ZM_2 \times \ZM_2 \times \Sigma$. One lets this group act on $C^\ast \Gg_{\Ll_0}$ by ignoring the augmented part. However, $T$ and $P$ are always represented by anti-unitary operators, hence the generators of the two $\ZM_2$ subgroups must act anti-linearly on the algebras of observables. This is effectively dealt with by passing to an extension of $\bar \Sigma$ by $\TM$ \cite{FreedAHP2013}. Up to isomorphisms, such extensions are enumerated by group morphisms $\phi : \bar \Sigma \to {\rm Out}(\TM) \simeq \ZM_2$ and by two-cocycles $\xi \in H^2_\phi(\bar \Sigma, \TM)$. We denote a group extension with such data by $\bar \Sigma^\phi_\xi$. As a set, $\bar \Sigma^\phi_\xi = \TM \times \bar \Sigma$ with multiplication determined by $\phi$ and $\tau$. The twist $\phi$ fixes which group elements will act anti-linearly and the twisting cocycle $\xi$ specifies whether $T$ or $P$ square to $1$ or $-1$. In addition, the elements of $\bar \Sigma$ are graded by a homomorphism $c : \bar \Sigma \to \ZM_2$, which assigns $c(P)=1$, $c(T)=0$ and $c(\sigma)=0$ for all $\sigma\in \Sigma$, to force the particle-hole exchange to be a graded symmetry that reverses the sign of a symmetric Hamiltonian.

The next task is to endow the algebra $M_\infty \otimes C^\ast \Gg_{\Xi_{\Ll_0}}$ of physical Hamiltonians with a natural $c$-graded action of this extension group. On $C^\ast \Gg_{\Xi_{\Ll_0}}$, $\bar \Sigma_\xi^\phi$ acts as
\[
(t,\bar \sigma) \cdot f = \begin{cases}
t \, \overline{\alpha_{\bar \sigma}(f)} & \text{if }\phi(\bar \sigma)=1\\
t \,\alpha_{\bar \sigma}(f) & \text{if }\phi(\bar \sigma)=0.
\end{cases}
\]
As for $M_\infty$, we recall that it can be endowed with a canonical action of any group $G$ with countable set of irreducible representations (irreps) that are all finite dimensional. Indeed, by considering the universal representation Hilbert space 
\[
\Hh_G =  \bigoplus_{\rm irreps} V_{\rm irrep} \otimes \ell^2(\NM),
\]
we can present $M_\infty$ as the algebra of compact operators over $\Hh_G$. The latter algebra accepts an action $\beta$ by $G$ which is canonically induced from the action of $G$ on $\Hh_G$. It is uniquely determined by the requirement $\pi(g)(Kv)= \beta_g(K)\pi_g(v)$ for any $g\in G$ and $v \in \Hh_G$. When a grading $c$ is present, we restrict the above sum to the $c$-graded representations. Applying these elements to $\bar \Sigma^\phi_\xi$, we can and we do endow the algebra of physical Hamiltonians with the required action.

We now focus on the setting of this paper where $\Sigma = \langle \sigma_r \rangle$, and where time-reversal and particle-hole symmetries are broken, but chiral symmetry $S=TP$ is not. Thus, we have the symmetry group $\Gamma =\langle S, \sigma_r \rangle \subset \overline{\Sigma}$. As a combination of two anti-unitary transformations, $S$ is unitary, hence the data $(\phi,\tau)$ is trivial, but the grading $c(S) = 1$ is odd. All $c$-graded representations of $\Gamma$ are of the form $M_2 \otimes L(V)$, where $S$ is implemented by conjugation with ${\rm diag}(1,-1) \otimes I$, and $V$ is a representation space for $\Sigma$. As such, the symmetric Hamiltonians are encoded by self-adjoint elements 
\begin{equation}\label{Eq:ChiralH}
\begin{pmatrix}
    0 & W^\ast \\ W & 0 
\end{pmatrix}, \quad W \in M_\infty \otimes C^\ast \Gg_{\Xi_{\Ll_0}}
\end{equation}
where $M_\infty$, hence $M_\infty \otimes  C^\ast \Gg_{\Xi_{\Ll_0}}$ too, are endowed with the canonical action by $\Sigma$. Then $W$ must be fixed by this action to have a fully symmetric Hamiltonian.

We now inspect the action of $\Sigma$ on the transversal $\Xi_{\Ll_0}$ displayed in Fig.~\ref{Fig:Transversal}. It is simply given by reflection of the diagram w.r.t. the main diagonal. What is important is the existence of closed subsets invariant to the groupoid and $\Sigma$ actions. They are $\Xi_2 = \Xi_{\Ll_0}$, {\it i.e.} the whole phason space, $\Xi_1$, and $\Xi_0 = \{\Ll_b\}$, {\it i.e.} the bulk lattice, both shown in Fig.~\ref{Fig:Filtration}. They generate a maximal filtration of the phason space
\begin{equation}\label{Eq:Filtration}
    \{\Ll_b\}=\Xi_0 \subset \Xi_1 \subset \Xi_2= \Xi_{\Ll_0}
\end{equation}
by invariant closed subsets. As noted in \cite{OjitoCMP2025} and emphasized in again in our introduction, such canonical structures of the phason space are key to the operator theory approach to HOTIs. 

\subsection{Defect algebras, gappable edges, and all that}\label{Sec:DefectAlgebras}

\begin{figure}[t!]
\center
	\includegraphics[width=0.9\linewidth]{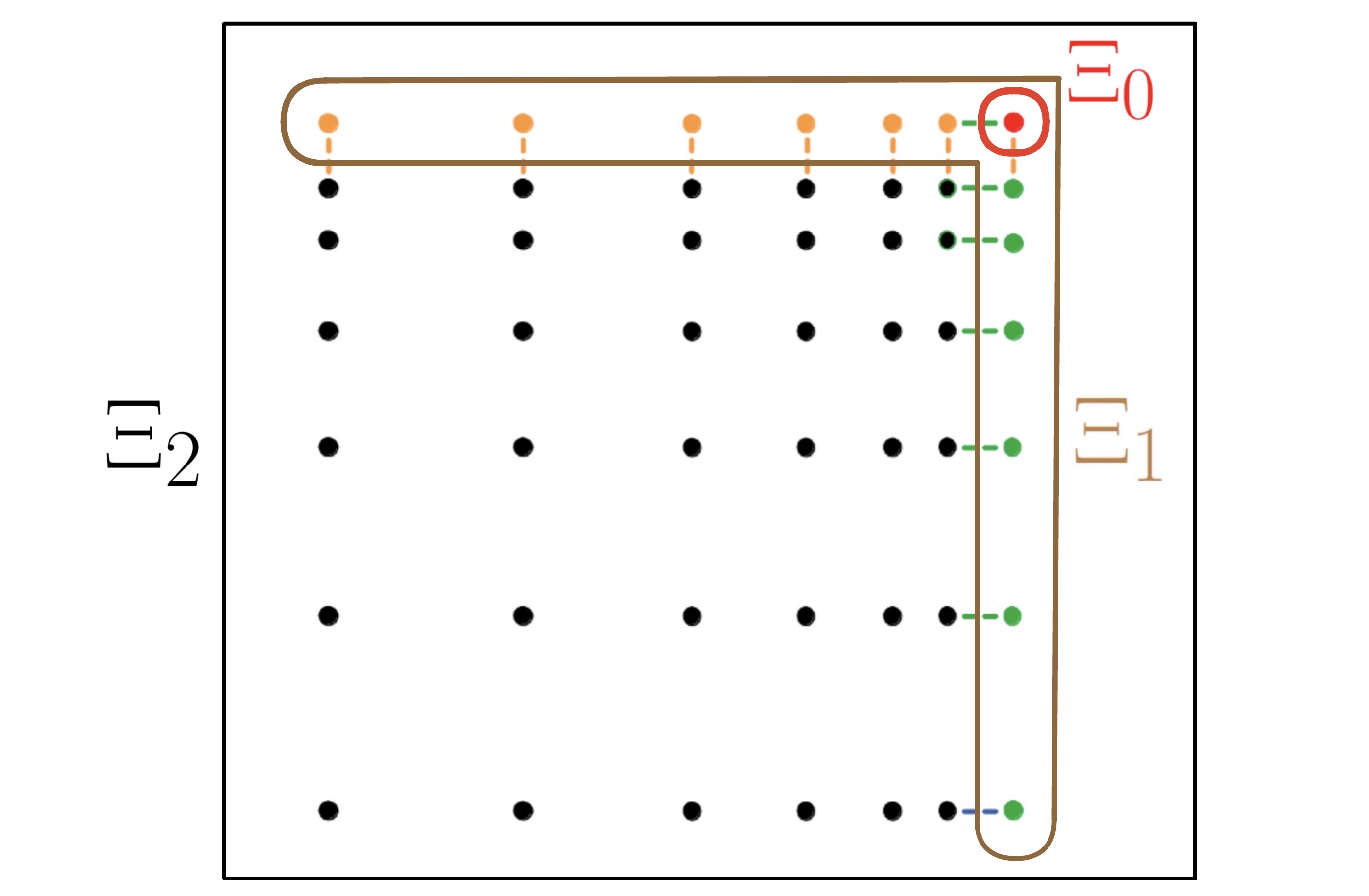}
	\caption{\small Maximal filtration of the phason space.}
	\label{Fig:Filtration}
\end{figure}

Given a closed subset $\Xi$ of the phason space $\Xi_{\Ll_0}$, invariant against the groupoid's action, we can make sense of the groupoid $\Gg_\Xi$ by simply restricting the triples~\eqref{Eq:Triples} to those with $\Ll \in \Xi$. Thus, for each set $\Xi_i$ in \eqref{Eq:Filtration}, we have the $C^\ast$-algebra $C^\ast \Gg_{\Xi_i}$ and, since we insisted that each $\Xi_i$ be $\Sigma$-symmetric, these algebras come each with a canonical graded action of desired symmetries. Furthermore, they enter the relation
\begin{equation}\label{Eq:Cofiltration}
    C^\ast \Gg_{\Xi_2} \stackrel{\mathfrak p^{2}}{\twoheadrightarrow} C^\ast \Gg_{\Xi_1} \stackrel{\mathfrak p^{1}}{\twoheadrightarrow} C^\ast \Gg_{\Xi_0},
\end{equation}
where the surjective maps $\mathfrak p^i$ are $\Sigma$-equivariant and simply given by restrictions of functions defined over $\Gg_{\Xi_i}$ to the smaller $\Gg_{\Xi_{i-1}}$. The above relation amplifies without problem when we take the tensor product with $M_2 \otimes M_\infty$ so, to simplify the next discussion, we momentarily leave this tensoring aside. 

The kernel of $C^\ast \Gg_{\Xi_i} \twoheadrightarrow C^\ast \Gg_{\Xi_{i-1}}$ is simply $C^\ast \Gg_{\Xi_i \setminus \Xi_{i-1}}$ and, by examining which $\Ll$'s are contained in $\Xi_i \setminus \Xi_{i-1}$ in Fig.~\ref{Fig:Filtration}, we see that $\Xi_2 \setminus \Xi_1$ contains the black dots, while $\Xi_1 \setminus \Xi_0$ contains the green and orange dots, excluding the accumulation red point. In other words, if $\Ll \in \Xi_2 \setminus \Xi_1$, then the observer is examining the dynamics with the corner inside of its field of view. We want to explain how algebra $C^\ast \Gg_{\Xi_2 \setminus \Xi_1}$ generates operators that are confined near the corner. To start, we restrict the discussion to the dense subalgebra $\CM \Gg_{\Xi_2 \setminus \Xi_1}$, which contains the functions with compact support. Now, $\Xi_2 \setminus \Xi_1$ is a discrete set, so its compact subsets are actually finite subsets. Similarly, any $\Ll$ is a discrete subset so again its compact subsets are finite subsets. As a result, a function over $\Xi_2 \setminus \Xi_1$ with compact support will have non-zero values only for a finite number of triples engaging a finite number of black dots in Fig.~\ref{Fig:Filtration}. Lastly, by examining the expression of the representations \eqref{Eq:PL}, we see that, indeed, all matrix elements there are confined in a finite region which has the corner in its field of view. The passage to the $C^\ast$ algebra relaxes this condition but still imposes a localization of the operators. Indeed, the operators from the $C^\ast$-algebra can all be approximated by operator from $\CM \Gg_{\Xi_2 \setminus \Xi_1}$, which assures us that their matrix elements fade away as we move away from the corner. In conclusion, we can call $C^\ast \Gg_{\Xi_2 \setminus \Xi_1}$ the corner algebra. 

\begin{figure}[t!]
\center
	\includegraphics[width=0.9\linewidth]{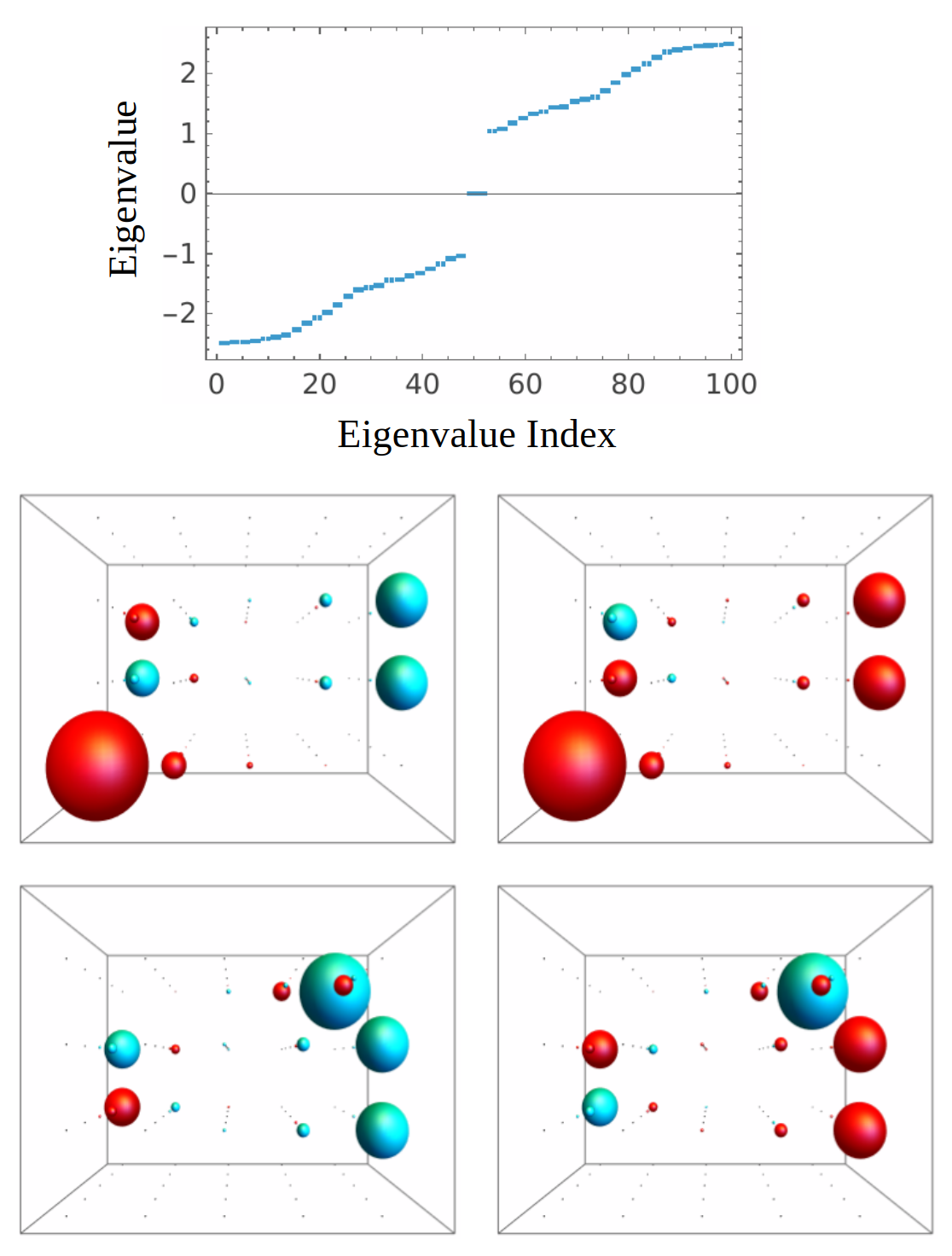}
	\caption{\small Spectrum and renderings of the zero modes of the tight-binding Hamiltonian \eqref{Eq:TightBindingH}, as computed on a $5 \times 5$ lattice with open boundary conditions. The vertical degrees of freedom represent the four orbital indices. The site-amplitudes of the modes are encoded in the sizes of the spheres, while colors encode their signs, with red for negative and teal for positive.}
	\label{Fig:TBP}
\end{figure}

Similarly, if $\Ll \in \Xi_1 \setminus \Xi_0$, then the observer is examining the dynamics with one of the edges in its field of view but the corner is beyond the horizon. By the same mechanism explained above, the algebra $C^\ast \Gg_{\Xi_1 \setminus \Xi_0}$ produces operators with matrix elements localized around the edges, hence it should be called the edge algebra.

Now, let $h \in M_2 \otimes M_\infty \otimes C^\ast \Gg_{\Xi_{\Ll_0}}$ be a self-adjoint element encoding a dynamics of the system. As we already emphasized, this dynamics can be observed from many points in space and what will be reported from there is fully encoded in the representation $\pi_\Ll(h)$ (see~\eqref{Eq:PL}), where $\Ll \in \Xi_{\Ll_0}$ is the lattice seen by the observer. As a result, we can translate the intuitive notions of gapped and gappable dynamics at the edges into very precise statements:

\begin{definition}
    The dynamics is said to be
    \begin{itemize}
    \item Gapped in the bulk if $\pi_{\Ll_b}(h)$ is a gapped operator.
    \item Gapped at the edges if the operator $\pi_\Ll(h)$ is gapped for all $\Ll \in \Xi_1 \setminus \Xi_0$. 
    \item Gappable at the edges while gapped in the bulk if $h$ admits an equivariant homotopy inside $M_2 \otimes M_\infty \otimes C^\ast \Gg_{\Xi_{\Ll_0}}$ that stays gapped in the bulk and the end result is gapped at the edges.
    \end{itemize}
\end{definition}

Note that this definition makes statements about a dynamics in the presence of a corner, but how can we declare a bulk Hamiltonian as being gapped or gappable at the edges? Here is one of many places where the maps in \eqref{Eq:Cofiltration} will come handy. Note that the map ${\rm id} \otimes \mathfrak p^2$ associates to any $h$ from $M_2 \otimes M_\infty \otimes C^\ast \Gg_{\Xi_{\Ll_0}}$ a unique bulk-edge Hamiltonian $h_{be} = ({\rm id} \otimes \mathfrak p^2) (h) \in M_2 \otimes M_\infty \otimes C^\ast \Gg_{\Xi_1}$, and this bulk-edge Hamiltonian is further mapped by ${\rm id}\otimes \mathfrak p^1$ to a unique bulk Hamiltonian $h_b = ({\rm id} \otimes \mathfrak p^1)(h_{be}) \in M_2 \otimes M_\infty \otimes C^\ast \Gg_{\Xi_0}$. Because the maps $\mathfrak p^i$ are surjective, any bulk Hamiltonian can be produced by either projecting all the way from $M_2 \otimes M_\infty \otimes C^\ast \Gg_{\Xi_{\Ll_0}}$ or from $M_2 \otimes M_\infty \otimes C^\ast \Gg_{\Xi_1}$. Since $({\id} \otimes \pi_\Ll)(h)$ being gapped for all $\Ll \in \Xi_1$ is equivalent to $({\rm id} \otimes \mathfrak p^1)(h)$ being gapped as an element of the algebra $M_2 \otimes M_\infty \otimes C^\ast \Gg_{\Xi_1}$, the following definition of gappable bulk Hamiltonian writes by itself:
\begin{definition}\label{Def:Gap2}
    A gapped bulk Hamiltonia $h_b \in M_2 \otimes M_\infty \otimes C^\ast \Gg_{\Xi_0}$ is gappable at the edges if we can find a gapped equivariant homotopy $h_b(s)$ inside $M_2 \otimes M_\infty \otimes C^\ast \Gg_{\Xi_0}$ such that $h_b(1)$ has a pre-image in $M_2 \otimes M_\infty \otimes C^\ast \Gg_{\Xi_1}$, under ${\rm id} \otimes \mathfrak p^1$ which is gapped. 
\end{definition}

 Since the difference between two such pre-images belongs to the kernel of ${\rm id} \otimes \mathfrak p^1$, hence to the edge algebra $M_2 \otimes M_\infty \otimes C^\ast \Gg_{\Xi_1 \setminus \Xi_0}$, the above statement effectively says that we can adiabatically and equivariantly deform $h_b$ until we are presented with the opportunity of adding an edge potential that gaps the dynamics at the edges.

\subsection{The intrinsic bulk-corner correspondence}\label{Sec:BulkCorner}

We will engage $\Sigma$-equivariant $K$-theories of the $C^\ast$ algebras we already introduced. Given a generic $C^\ast$-algebra $\Aa$ equipped with a $\Sigma$-action, the $K_0^\Sigma$ and $K_1^\Sigma$ theories that we need can be presented in two equivalent ways, which both are useful in the present context (see \cite{ProdanRMP2016} for a detailed account). In one description, an entry of $K_0^\Sigma(\Aa)$ is a set containing all gapped self-adjoint $\Sigma$-symmetric elements from $M_\infty \otimes \Aa$ that can be connected by gapped homotopies, while an entry of $K_1^\Sigma(\Aa)$ is a set containing all gapped self-adjoint $\Sigma$-symmetric odd-graded elements from $M_2 \otimes M_\infty \otimes \Aa$ that can be connected by gapped homotopies. Thus any gapped $\Sigma$ and chiral symmetric Hamiltonian $h$ defines a $K_1^\Sigma$-class $[h]_1$ and, when topological class of a $\Sigma$ and chiral symmetric Hamiltonian is mentioned in a discussion, the reference is really about its $K_1^\Sigma$-class. 

It should be noted that, in equivariant $K$-theory, the algebra is automatically stabilized by tensoring with $M_\infty$ or $M_2 \otimes M_\infty$, which are both endowed with the canonical group action. This allows for a natural addition operation of classes by stacking: If $h$ and $h'$ are gapped Hamiltonians from $M_2 \otimes M_N \otimes \Aa$ and $M_2 \otimes M_{N'}\otimes \Aa$, respectively, then $[h]_1 \oplus [h']_1 = [h\oplus h']_1$, where $h+h'$ is seen as an element of $M_2\otimes M_{N+N'} \otimes \Aa$. Furthermore, it should become clear that HOTI phenomena that are stable against switching on and off orbitals emerge entirely from the groupoid algebras. The orbital spaces' only role is to accommodate the generators of the $K$-theories.

We want to show now how $K$-theory answers challenge (2) stated in our introduction. For this, let us recall the simple fact that if $\mathfrak p : \Aa \to \Bb$ is a unital morphism of $C^\ast$-algebras, then gapped self-adjoint elements are mapped into gapped self-adjoint elements. As such, we have a well defined maps $\mathfrak p_i$ at the level of $K$-theories, $[h]_i \mapsto [\mathfrak p_i(h)]_i$, $i=0,1$. Definition~\ref{Def:Gap2} can then be reformulated as follows:

\begin{definition}
    A symmetric bulk Hamiltonian $h_b$ is gappable at the edges if $[h_b]_1 \in K_1^\Sigma(C^\ast \Gg_{\Xi_0})$ accepts a pre-image under the map $\mathfrak p^1_1 : K_1^\Sigma(C^\ast \Gg_{\Xi_1}) \to K_1^\Sigma(C^\ast \Gg_{\Xi_0})$.
\end{definition}

We are now in the position to identify all symmetric Hamiltonians gappable at the edges. Indeed, according to this definition, which embodies the intuitive picture and nothing else, all symmetric bulk Hamiltonians gappable at the edges are contained in the $K$-theoretic classes found inside the image ${\rm Im} \, {\mathfrak p^1_1} \subset K_1^\Sigma(C^\ast \Gg_{\Xi_0})$.

In the second picture of $K$-theories, the entries of $K_0^\Sigma$ are classes of stably homotopic $\Sigma$-symmetric projections, while the entries of $K_1^\Sigma$-group are classes 
of stably homotopic symmetric unitaries. This picture is useful because, in order to identify the topological defect modes, we are not looking for a Hamiltonian but rather for either projections or unitaries that encode or expose them. $K$-theory offers a machinery to do just that, which we now explain for the present context. 

Consider a surjective equivariant homomorphism $\mathfrak p : \Aa \twoheadrightarrow \Bb$. Then there exist canonical maps in $K$-theories 
\begin{equation}\label{Eq:BoundaryM}
\partial_i : K^\Sigma_i (\Bb) \to K_{i+1}^\Sigma({\rm Ker} \, \mathfrak p), \quad i=0,1,
\end{equation}
called boundary maps. If a gapped Hamiltonian from the stabilization of $\Bb$ lands in a non-trivial $K$-theoretic class of ${\rm Ker} \, \mathfrak p \subset \Aa$ under this map, we can be sure that it does not have a gapped pre-image in the stabilization of $\Aa$. Otherwise, we can be sure that it does. Now, if $\mathfrak p$ is $\mathfrak p^2$ in \eqref{Eq:Cofiltration}, its boundary map $\partial^2_1$ enables us to identify all gapped $\Sigma$ and chiral symmetric elements from $M_2 \otimes M_\infty \otimes C^\ast \Gg_{\Xi_1}$ that become gapped when lifted to $M_2 \otimes M_\infty \otimes C^\ast \Gg_{\Xi_{\Ll_0}}$, {\it i.e.} to the full quarter-plane geometry. Now, Definition~\ref{Def:Gap2} assures us that any symmetric bulk Hamiltonian gappable at the edges has at least one gapped symmetric pre-image in $M_2 \otimes M_\infty \otimes C^\ast \Gg_{\Xi_1}$, and these pre-images will produce corner modes when extended to quarter-plane geometry, provided they are mapped by $\partial_1^2$ into a nontrivial $K_0^\Sigma$-theoretic class of the corner algebra $C^\ast \Gg_{\Xi_2 \setminus \Xi_1}$. In such cases, those classes correspond to the difference of $\Sigma$-symmetric finite-rank projections $[P_0^+]_0 - [P_0^-]_0$, where $P_0 = P_0^+ + P_0^-$ is the projection onto the $0$-energy corner modes, as resolved by the chiral sectors (see Prop.~4.3.3 in \cite{ProdanSpringer2016} and comments in \cite{OjitoCMP2025}).  

As a last step, we need to address what happens when we impose different boundary conditions at the edges. This is related to the fact that there can be many pre-images of $[h_b]_1$, differentiated by the many symmmetry-compliant boundary conditions we can impose. These pre-images can be pushed by $\partial_1^2$ into different corner projections displaying different corner modes, but they should be counted as being the same under intrinsic HOTI paradigm. To do so, we note that the difference between two pre-images is always inside ${\rm Ker} \, \mathfrak p_1^1 \in K_1^1(C^\ast \Gg_{\Xi_1 \setminus \Xi_0})$ and, if we want to list the topological corner modes up to shuffling under different allowed boundary conditions, we need to quotient out $\partial_1^2 ({\rm Ker} \, \mathfrak p_1^1)$ from the range of $\partial_1^2$. This removes the ambiguities related to the pre-images, hence with the boundary conditions, and we obtain a well defined bulk-corner correspondence map give by the sequence, 
\[
\mbox{bulk-model} \to \mbox{pre-image} \to {\rm Im} \, \partial_1^2 \to \frac{{\rm Im} \, \partial_1^2}{\partial_1^2({\rm Ker}\,\mathfrak{p}^1_*)},
\]
which answers challenge (3) stated in our introduction:
\begin{proposition}[\cite{OjitoCMP2025}]\label{Prop:BC1}
There is a map 
\begin{equation}\label{Eq:BC}
\delta_{{BC}}\colon {\rm Ker} \, \mathfrak p_1^1 \longrightarrow \frac{{\rm Im} \, \partial_1^2}{\partial_1^2({\rm Ker}\,\mathfrak{p}^1_*)} 
\end{equation}
whose image identifies all intrinsic topological corner modes generated by symmetric gappable models. Furthermore, a symmetric bulk model is an intrinsic HOTI if and only if its $K_1^\Sigma$-class belongs to the domain of $\delta_{BC}$ but is outside of ${\rm Ker}\, \delta_{BC}$. 
\end{proposition}

\begin{figure}[t!]
\center
	\includegraphics[width=0.8\linewidth]{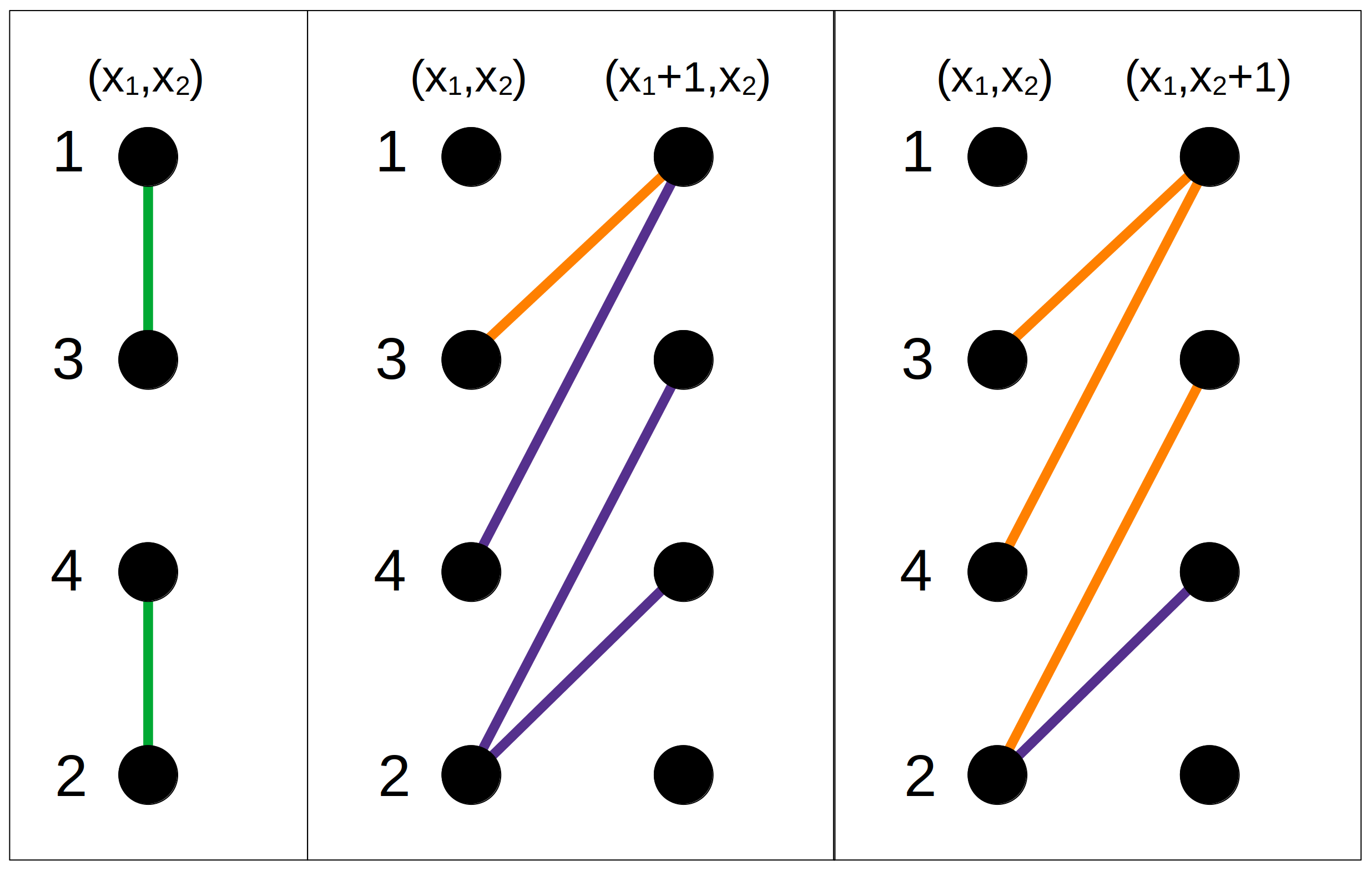}
	\caption{\small Coupling diagram of  the acoustic system as derived from \eqref{Eq:TightBindingH}, with the acoustic cavities shown as black disks. Since the model has four orbitals per atom, the acoustic system will consist of four 2-dimensional layers, each of which representing one orbital (see the orbital indices displayed in each panel). The left panel displays the mode-couplings at the same site. The middle (right) panel displays the mode-couplings between neighboring sites in the $x_1$ ($x_2$) directions. Green bonds indicate couplings of strength $1/2$. Orange bonds represent negative while indigo bonds indicate positive couplings of strength 1.}
	\label{Fig:Couplings}
\end{figure}

\begin{figure}[t!]
\center
	\includegraphics[width=0.99\linewidth]{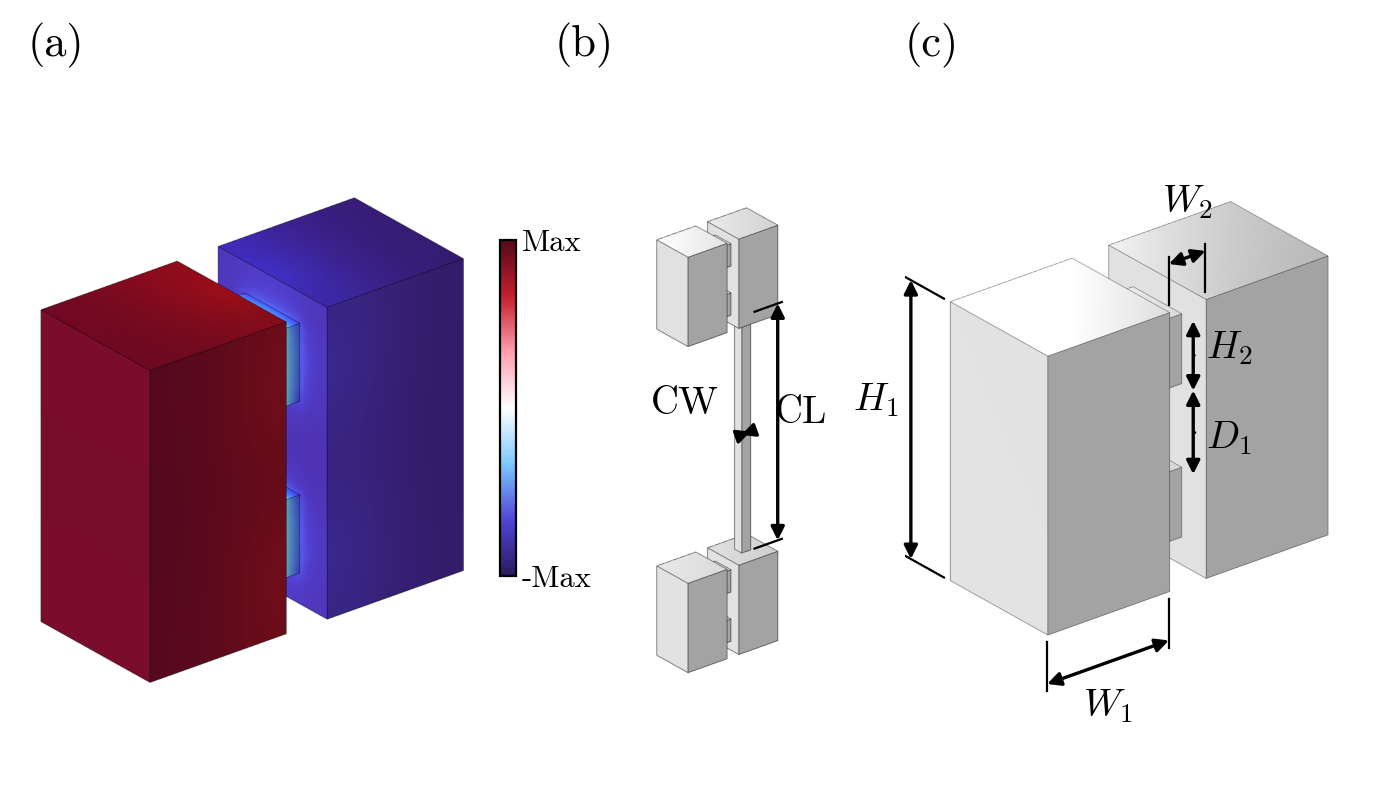}
	\caption{\small The H-resonators and their coupling. a) Acoustic pressure profile of the chamber-antisymmetric mode of a single resonator at $f_0 = 4.85$ kHz, which serves as the base mode prior to coupling. The two chambers oscillate in antiphase with nearly uniform pressure, so the sign of a coupling is set by the chamber on which a bridge lands. b) COMSOL geometry of two coupled resonators, with the coupling length (CL) and the coupling width (CW) labeled. c) Relevant dimensions of the double-bridged H-resonator: $H_1 = 20$~mm, $W_1 = 10$~mm (each of the two large rectangular prisms is $H_1\times W_1 \times W_1$), $W_2 = 3$~mm, $H_2 = 5$~mm (each bridge is an $H_2\times H_2\times W_2$ rectangular prism centered on $W_1$), $D_1 = 6$~mm.}
	\label{Fig:HResonator}
\end{figure}

This can be formulated in a far more streamlined language, which we reproduce here from \cite{OjitoCMP2025}  for the experts, without explaining it (see \cite{OjitoCMP2025} for details). The cofiltration~\eqref{Eq:Cofiltration} gives rise to a spectral sequence $(E^{r}_{p,q}, d^{\,r}_{p,q})$ converging to the equivariant $K$-theories of $C^\ast \Gg_{\Xi_{\Ll_0}}$.  
Then the bulk–corner correspondence map, defined above, coincides with \(d^{\,2}_{0,1}\), which has exactly the same domain and target space as in~\eqref{Eq:BC}. Thus, despite its highly specialized form, the bulk-corner correspondence map can be identified with standard objects in topology. This picture accepts an extremely broad generalization (see Thm.~1.4 and examples in \cite{OjitoCMP2025}).

\begin{figure*}[t!]
\center
	\includegraphics[width=0.8\linewidth]{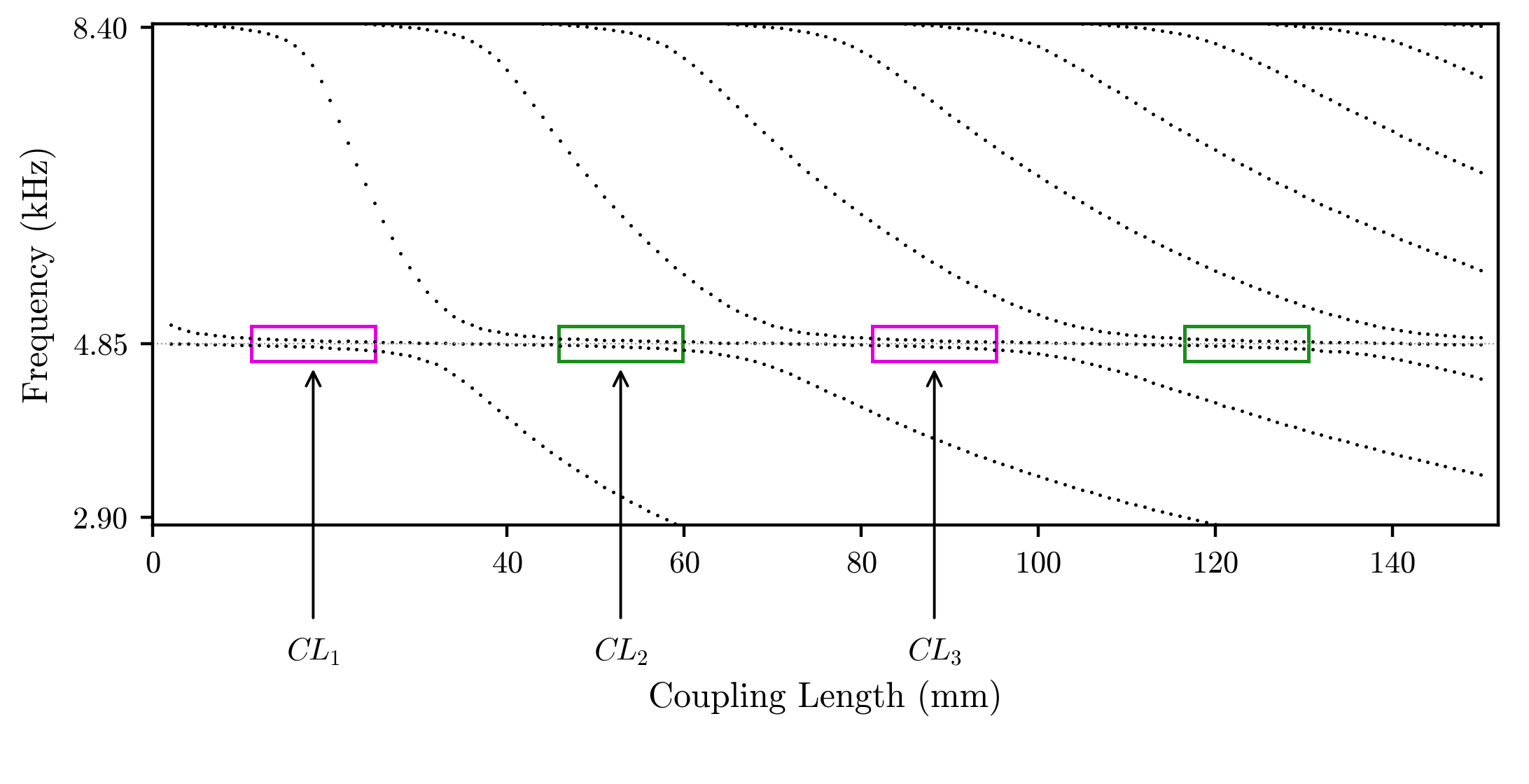}
	\caption{\small Dimer's resonant spectrum as a function of bridge length, for a bridge width of 2.3~mm. Boxes mark the windows where the two resonator modes sit symmetrically about $f_0$, so that the bridge implements a pure hopping. They occur at odd multiples of a quarter wavelength, $CL \approx 18$, 53, 88 and 124~mm, and consecutive windows give couplings of opposite sign (green and purple boxes). The arrows mark the three windows used, $CL_1, CL_2, CL_3$. In the lattice, where the bridges bend around the resonators, their lengths are $CL_1 = 16.5\,\mathrm{mm}$ (on-site couplings), $CL_2 = 54$--$55\,\mathrm{mm}$ and $CL_3 = 89$--$91\,\mathrm{mm}$. While only three distinct couplings are required for this implementation, the spectrum can be extended to accommodate additional coupling values.}
	\label{Fig:DimerVsLength}
\end{figure*}

\subsection{Identifying the HOTIs}

For the corner algebra, we have 
\[
K_0^\Sigma(C^\ast \Gg_{\Xi_2 \setminus \Xi_1}) = \Z[ \chi_+\otimes E_0]_0 \oplus \Z[\chi_-\otimes E_0]_0,
\]
where $E_0 = |0\rangle \langle 0 \rangle$ is the rank-one projection onto the corner site, and $\chi_\pm$ denote the central projections supporting the trivial and non-trivial irreducible representations of $\Sigma \simeq \Z_2$ on $\C^2$. Furthermore, for the bulk algebra,  
$$K_1^\Sigma(C^\ast \Gg_{\Xi_0}) = \ZM[U]_1 \oplus \ZM[W]_1,$$  where the unitaries are
\[
U=\chi_+\otimes S_1S_2+ \chi_-\otimes {\bf 1}
\]
and
\begin{equation}\label{Eq:GenW}
W=\tfrac{1}{2}\begin{pmatrix}
        -S_1^\ast-S_2^\ast & S_1^\ast-S_2^\ast\\
        S_1-S_2 & S_1+S_2
    \end{pmatrix}
\end{equation}
with the latter written as an element of $M_2(\CM) \otimes C^\ast \ZM^2$ with the mirror action on the orbitals given by $M  = {\scriptsize \begin{pmatrix} 1 & 0 \\ 0 & -1 \end{pmatrix}}$, such that $(Ad_M \otimes \alpha_{\sigma_r})(W) = W$. To get a geometric description of the generators, observe first that $C^\ast \Gg_{\Xi_0} \simeq C^\ast \ZM^2 \simeq C(\mathbb{T}^2)$. Viewing $\mathbb{T}^2$ as the square with opposite edges identified, the unitary $U$ has winding number $1$ along each pair of opposite edges and $W$ becomes diagonal matrix when restricted to the diagonal of the square; in that restriction, its components have winding numbers $1$ and $-1$ in the eigenspaces of $\chi_+$ and $\chi_-$, respectively. From this observation, it follows  that ${\rm Im} \, \mathfrak p_1^1 = \mathbb{Z}[W]_1$, thus the gappable models are all stacks of an elementary Hamiltonian obtained by inserting $W$ into \eqref{Eq:ChiralH}. 

Now, a computation shows that
$$
\partial_1^2({\rm Im} \, \mathfrak p_1^1) = \ZM[\chi_+\otimes E_0]_0,
$$
while 
$$
\partial_1^{\,2}({\rm Ker} \, \mathfrak{p}^1_1)) = 
2 \ZM [\chi_+\otimes E_0 ]_0,
$$
from where we conclude that, 
$$
{\rm Im}\, \delta_{BC}
= \mathbb{Z}_2 [\chi_+\otimes E_0 ]_0. 
$$
One can summarize all this information as:

\begin{theorem}\label{theo: main}
The symmetric bulk Hamiltonian $H$ derived from $W$ exhibits non-trivial corner modes whose parity is protected by the bulk topology and the symmetry $\Sigma$. Moreover, any other symmetric Hamiltonian with same characteristics is stably homotopic with odd number of stakings of $H$.
\end{theorem}

\section{Mirror winding number}
We present a practical method to decide if a symmetric bulk model gappable at the edges is a HOTI. We will work in the momentum space and identify the bulk algebra with $C(\mathbb{T}^2)$, where $\mathbb{T}^2$ is regarded as the square $[0,1]^2$ with opposite edges identified. Denote by $D$ the diagonal of $\mathbb{T}^2$, which is a closed $\Sigma$-invariant subset of $\T^2$. Then the restriction map $
\mathfrak{d}\colon C(\TM^2)\to C(D)$
defines a surjective $\Sigma$-equivariant $\ast$-homomorphism. Consequently, $\mathfrak{d}$ induces maps in the equivariant $K$-theory $\mathfrak{d}_i\colon K_i^\Sigma(C(\TM^2))\to K_i^\Sigma(C(D))$. One can easily check $K_0^\Sigma(C(D))\simeq \mathbb{Z}^2\simeq K_1^\Sigma(C(D))$ by noticing that  $D\simeq \SM^1$ and $\Sigma$ acts trivially on $D$. The generators of $K_1^\Sigma(C(D))$ can be identified with $u=\chi_+\otimes u_D+\chi_-\otimes {\bf 1}_D$ and $w=\chi_+\otimes {\bf 1}_D +\chi_-\otimes u_D $, where $u_D:D\to \SM^1$ is any function with winding number $1$. Therefore, $\mathfrak{d}_1([U]_1)=2[u]_1$ and $\mathfrak{d}_1([W]_1)=[u^*w]_1$.

\begin{figure*}[t!]
\center
	\includegraphics[width=0.99\linewidth]{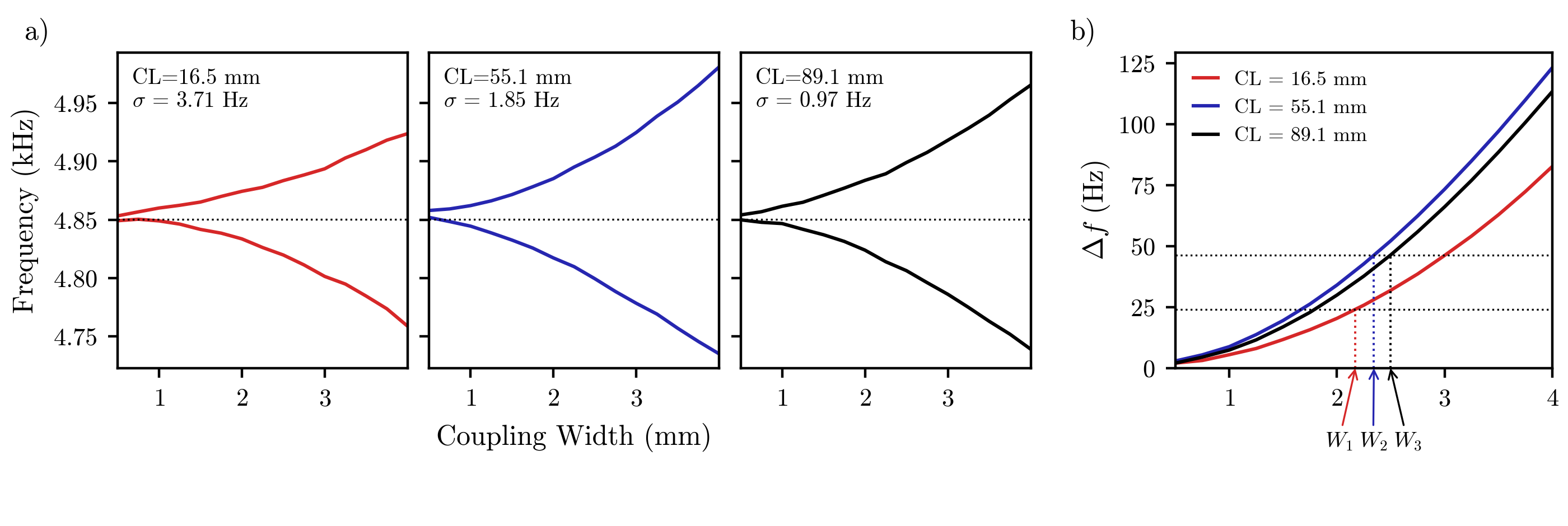}
	\caption{\small Hybridization diagrams. a) Dimer's resonant frequencies as function of bridge width for the three chosen coupling lengths (CLs), computed with the bridge geometries used in the lattice ($CL = 16.5$, 55.1 and 89.1~mm). Parameter $\sigma$ denotes the standard deviation of the average frequency from $f_0$, indicated by the dotted line ($\sigma = 3.71$, 1.85 and 0.97~Hz), confirming the high symmetry of the hybridization diagrams. b) Coupling strength $\Delta f$ of the three CLs in a). The intersections with the target strengths (dotted lines: 24.1~Hz for strength $1/2$ and 46.3~Hz for strength 1) determine the widths $W_1 = 2.17$, $W_2 = 2.34$ and $W_3 = 2.50$~mm that yield equivalent coupling strengths across different CL values.}
	\label{Fig:Hybridization}
\end{figure*}

Let now $h_b\in  M_2 \otimes {\rm M}_N(\mathbb{C})\otimes C(\TM^2)$ be a gapped Hamiltonian that is $\Sigma$ and chiral symmetric, and let the unitary $U_{h_b}$ be the off-diagonal part of ${\rm sign}(h_b)$. Then $\mathfrak{d}(U_{h_b})\in {\rm M}_{N}(\C)\otimes C(D)$ is also a unitary so one may define $\nu_\pm (\mathfrak{d}(U_{h_b}))$ as the winding number of $\mathfrak{d}(U_{h_b})$ in the eigenspace of $\chi_\pm,$ respectively.

\begin{definition}
The \emph{mirror winding number} of $h_b$ is the integer number
\begin{equation}\label{eq:mirror-winding}
\nu(h_b)\;=\;\frac{\nu_+(\mathfrak{d}(U_{h_b}))-\nu_-(\mathfrak{d}(U_{h_b}))}{2}.
\end{equation}
\end{definition}

\begin{theorem}
Let $h_b\in {\rm M}_N(\mathbb{C})\otimes C(\TM^2)$ be a symmetric bulk Hamiltonian as above, which in addition is gappable at the edges. Then $h_b$ is an intrinsic HOTI iff $\nu(h_b)$ is odd.
\end{theorem}
\begin{proof}
In $K_1^\Sigma(C(\TM^2)$, the class of $U_{h_b}$ decomposes as
\[
[U_{h_b}]_1 = n_1\,[U]_1 \oplus n_2\,[W]_1.
\]
Since $h_b$ is assumed gappable on the edges, $n_1=0$. Therefore $
\mathfrak{d}_1([U_{h_b}]_1)=n_2\,[u^*w]_1$ and one checks directly that the mirror winding number of the representative $u^*w$  is $1$. As such $\nu(h_b)=n_2$. This tells us that $h_b$ is stably equivalent with a stack of $\nu(h_b)$ elementary HOTIs identified in the previous section. Then the statement follows from Thm.~\ref{theo: main}.
\end{proof}

\section{Implementation of HOTI with acoustics}\label{Sec:Implementation}

We already identified a model representing the unique class of chiral HOTIs for the quarter-space geometry with mirror symmetry against the main diagonal. However, by construction, the model induced by the unitary  $W$ from \eqref{Eq:GenW} has flat bands because the Hamiltonian squares to the identity. We prefer a model for which waves have dispersion and, therefore, we will present here a slightly modified HOTI model and its implementation with acoustics. All simulations were conducted via COMSOL Multiphysics's acoustic module \cite{COMSOL}, where  the boundaries were treated as hard walls and the speed of sound was fixed at 343 m/s.

\subsection{The model}\label{sec:model}
We will work with the two-dimensional model
\begin{equation*}
H =
2\begin{pmatrix}
    0 & W+\tfrac{1}{4} I_{2}\\
    W^\ast+\tfrac{1}{4} I_{2} & 0
\end{pmatrix},
\end{equation*}
where $W$ is as in \eqref{Eq:GenW}. One can check that $W+sI_{2\times 2}$ is invertible for $s \in [0,\tfrac{1}{4}]$ and obviously remains $\Sigma$-symmetric, hence the seen modification preserves the topological class of the model. The factor in front is an energy rescaling and is irrelevant but practically convenient. The model is invariant to mirroring against the main diagonal, which switches $ S_1 \leftrightarrow S_2$ and acts on the orbital by conjugation with $I_{2 \times 2} \otimes \sigma_z$.

To reveal the coupling matrices, we write the Hamiltonian as in \eqref{Eq:HBulk}:
\begin{equation}\label{Eq:TightBindingH}
H=w_{(1,0)}\otimes S_1 + w_{(0,1)}\otimes S_2
+w_{(0,0)}\otimes {\bf 1}
+\text{h.c.},
\end{equation}
where $w_{(0,0)}=\tfrac{1}{2}\scriptsize{\begin{pmatrix} 0_2&I_2\\ 0_2&0_2 \end{pmatrix}}$ and
\[
w_{(1,0)}=
\scriptsize{\begin{pmatrix}
0&0&0&0\\
0&0&1&1\\
-1&0&0&0\\
1&0&0&0
\end{pmatrix}},\ w_{(0,1)}=
\scriptsize{\begin{pmatrix}
0&0&0&0\\
0&0&-1&1\\
-1&0&0&0\\
-1&0&0&0
\end{pmatrix}}.
\]
From these matrices, we can read the required couplings for the network of acoustic cavities and build the coupling diagram illustrated in Fig.~\ref{Fig:Couplings}. Note that the second and third internal degrees of freedom are interchanged to obtain a more compact layout. 

We report the spectrum of the Hamiltonian~\eqref{Eq:TightBindingH} in Fig.~\ref{Fig:TBP}, as computed on a finite $5 \times 5$ lattice with open (Dirichlet) boundary conditions (hence similar to our COMSOL model). There are four zero modes and their corresponding eigenvectors are displayed in the same figure. These are hybridizations of four corner modes, whose mix-up is un-avoidable on a finite lattice. The minimal model with open boundary conditions has an accidental additional symmetry and only two of the four zero modes are robust under boundary conditions that respect the mirror symmetry against the first diagonal.

\subsection{Methods and Design}

The coupling diagram seen in Fig.~\ref{Fig:Couplings} can be considered highly complex when compared with the typical acoustic crystals reported in the literature. Indeed, it requires as many as eight couplings per resonator, both positive and negative, and there are second-nearest neighbor couplings. To account for the large number of sign-varying couplings per resonator, we rely on the H-resonator \cite{saunders2025acousticmetamaterialspositivenegative}, which was introduced precisely for this purpose \cite{WuSciB2024}. Here we use a double-bridged configuration, shown in Fig.~\ref{Fig:HResonator}, operated in its chamber-antisymmetric mode at $f_0 = 4.85$~kHz, which will serve as the orbital of the synthetic atom. In this mode the pressure is nearly uniform inside each chamber and opposite in the two chambers, so a bridge attached to one chamber or to the other couples with opposite sign. The next mode of the same resonator (8.45~kHz) does not offer this freedom, because there the two chambers oscillate in phase. The resonator couplings will be implemented by acoustic bridges, as illustrated in Fig.~\ref{Fig:HResonator}(b) for a dimer.

\begin{figure}[t!]
\center
	\includegraphics[width=0.99\linewidth]{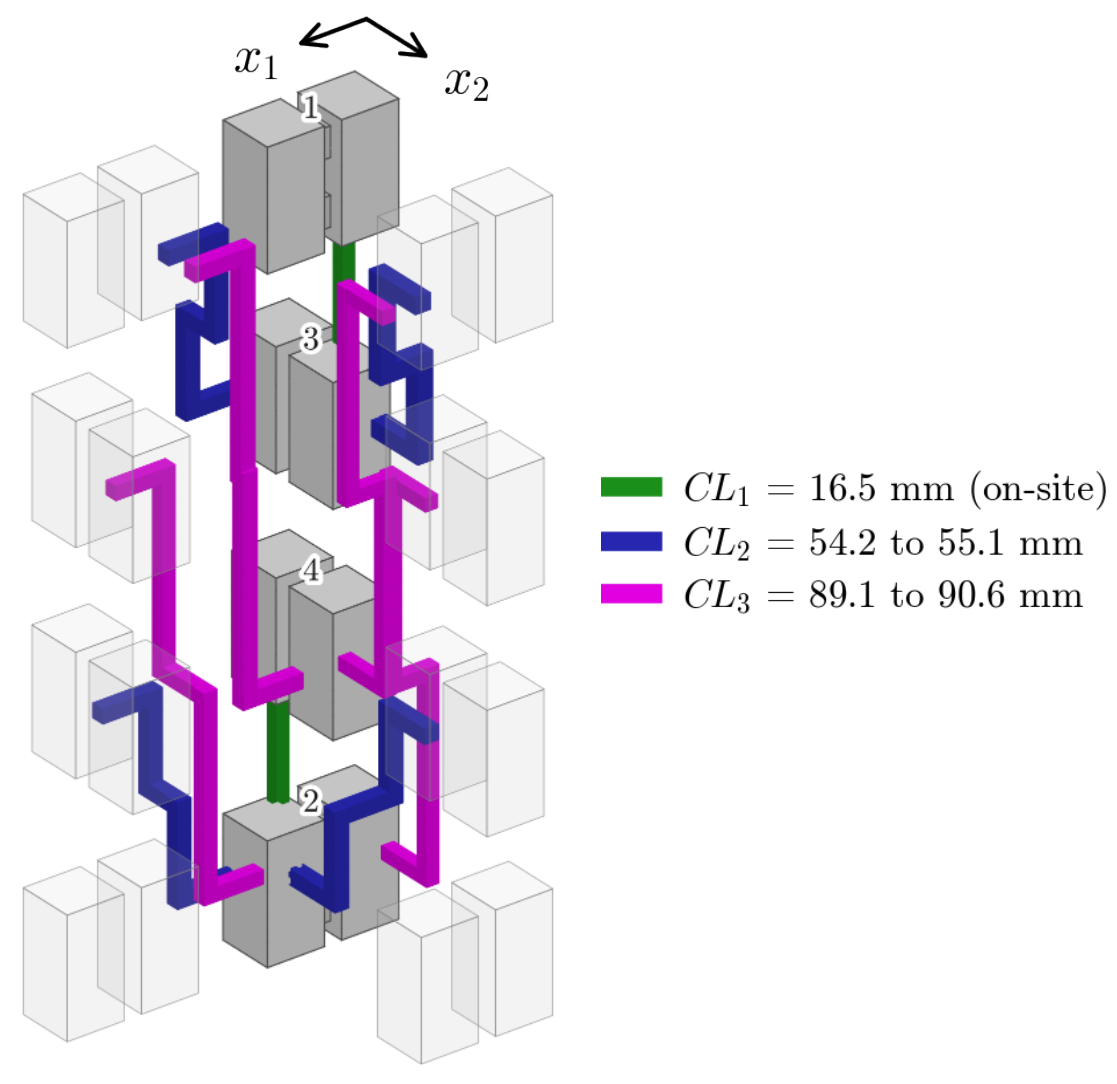}
	\caption{\small Repeating unit cell of the acoustic lattice. Four H-resonators are stacked in one column (orbitals 2, 4, 3, 1 from bottom to top); those carrying orbitals 1 and 2 are oriented along $x_1$, those carrying 3 and 4 along $x_2$. The two on-site bridges ($CL_1$, green) join the stacked resonators, and eight bridges of lengths $CL_2$ (blue) and $CL_3$ (magenta) connect the cell to its neighbors in the $x_1$ and $x_2$ directions (translucent). Bridge widths are 2.17--2.67~mm.}
	\label{Fig:UnitCell}
\end{figure}

The key to a model implementation, in general, is a reproducible map from bridge's geometry to tight-binding parameters. This is typically done via the so-called hybridization diagram, where the resonant frequencies of a dimer (and sometime trimer) are observed as function of one or more geometric parameters. In our case, it will be the cross-section of the acoustic bridges. Of course, if this cross-section is zero, there is no connection and we will observe two modes at exactly the frequency $f_0$ of the individual H-resonators. As the cross-section is increased, mode hybridization takes place and the frequencies split and drift. The frequencies and mode hybridizations are well characterized by a tight-binding model
\[
f_1 |1\rangle \langle 1| +f_2 |2\rangle \langle 2| +\Delta f (|1\rangle \langle 2| + |2\rangle \langle 1|),
\]
where $|i\rangle$ represent the already mentioned modes carried by the H-resonators of the dimer. We need to implement pure hopping terms, hence no on-site potentials, otherwise is almost impossible to enforce the chiral symmetry. For this, we must force $f_1$ and $f_2$ to stay at the $f_0$ value. Note that, in such cases, $\Delta f$ is genuinely equal to half of the frequency splitting observed in the hybridization diagrams.

Being able to eliminate on-site terms and control $\Delta f$ by varying the geometry of the bridges is our primary challenge, which is intensified by the complexity of the model. Indeed, as it can be already seen in the coupling diagram from Fig.~\ref{Fig:Couplings}, we are forced to work with bridges of three different lengths, and the one accommodating the second-near neighbor coupling will inherently exceed the wavelength of sound waves. This introduces a known challenge in meta-materials science, because the bridges themselves can support resonant modes that can strongly affect our control over the mode-couplings. Key to overcoming this challenge, is the observation that the resonances supported by the bridges move in frequency as their lengths are varied. As observed in \cite{WangPRL2023}, there are certain windows of lengths, which occur more or less periodically, where the bridge resonances stay away from the base frequency $f_0$ of the H-resonators and, as such, the coupling strengths induced by the bridges become predictable.

In Fig.~\ref{Fig:DimerVsLength}, we report the COMSOL simulated resonant frequencies of a dimer as function of bridge length, for a fixed bridge width of 2.3~mm. As one can see, there are wild swings in frequencies, except for well defined windows where the resonant frequencies occur close to $f_0$. It is in those windows where the mode-hybridization picture is valid. These windows occur at odd multiples of a quarter wavelength, $CL \approx (2m+1)\lambda/4$ with $\lambda = c/f_0 = 70.7$~mm, where a bridge does not shift the frequencies of the resonators it connects, and consecutive windows give couplings of opposite sign. The sign of every coupling is therefore fixed by two geometric choices: the chambers on which the bridge lands and the window of its length. Using this data, we identified the bridge lengths $CL_1 \approx 18$~mm, $CL_2 \approx 53$~mm and $CL_3 \approx 88$~mm. In the lattice, where the bridges must bend around the resonators, their lengths are 16.5~mm (on-site couplings), 54--55~mm and 89--91~mm.

\begin{figure*}[t!]
\center
	\includegraphics[width=0.99\linewidth]{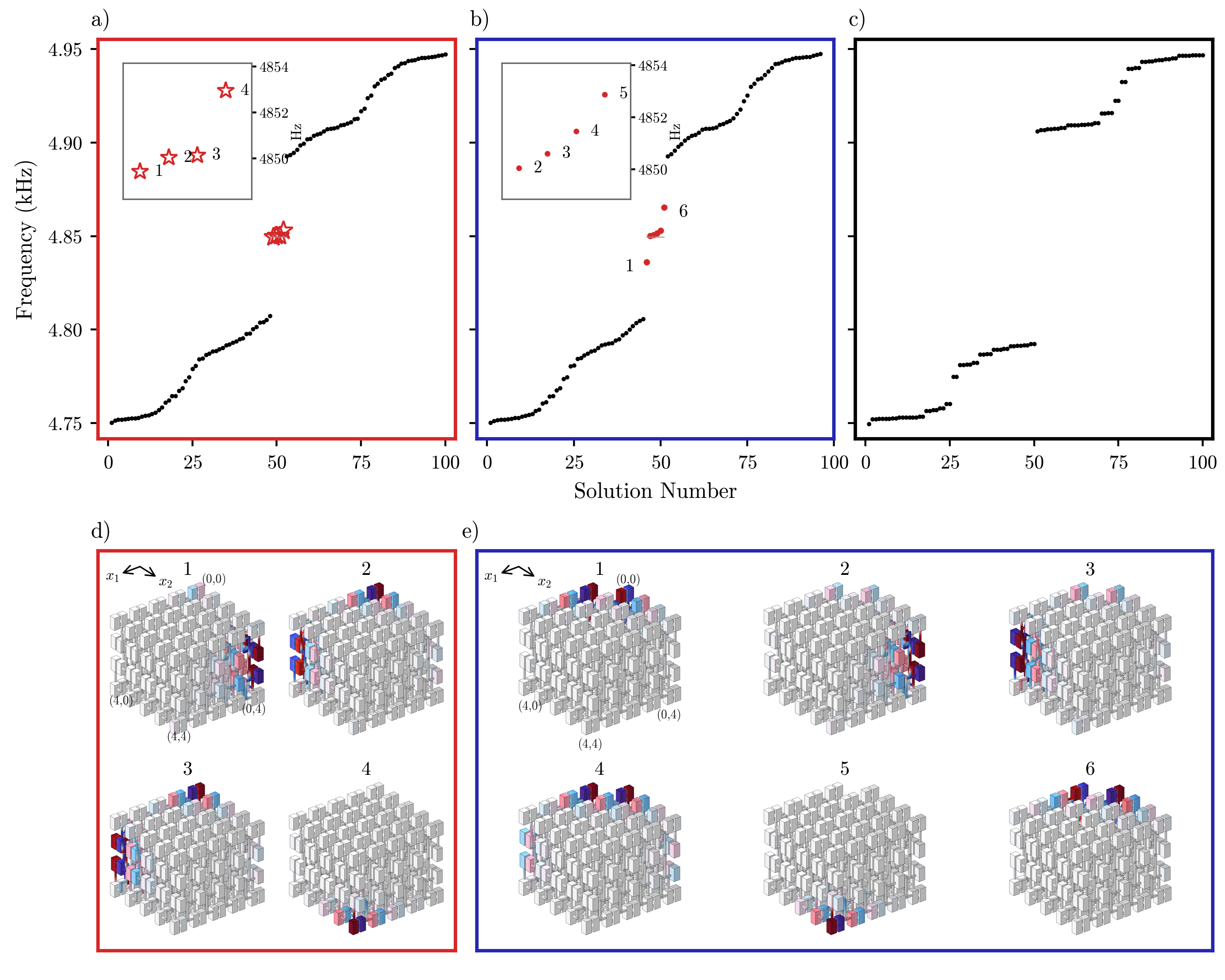}
	\caption{\small a) Band spectrum of the $5\times 5$ system shown in d) (100 resonator modes). Four modes lie inside the bulk gap (4807--4893~Hz), within 2.5~Hz of $f_0$ (inset), one at each corner: 1) $(0,4)$, orbitals 3/4; 2) and 3) the $(0,0)$ (orbital 1) and $(4,0)$ (orbitals 3/4) corner modes, which are degenerate to within 0.09~Hz and therefore appear hybridized; 4) $(4,4)$, orbital 2. b) Band spectrum of the system shown in e), where the corner cell $(0,0)$ is removed, a perturbation that preserves the mirror symmetry. Each in-gap mode is labeled according to its corresponding pressure profile shown in e). In-gap states: 2) $(0,4)$ and 3) $(4,0)$, orbitals 3/4, at 4850.0 and 4850.6~Hz; 4) the $(0,0)$ corner mode, which has moved onto orbital 1 of the cells $(0,1)$ and $(1,0)$, at 4851.5~Hz; 5) $(4,4)$, orbital 2, at 4852.9~Hz. States 1) and 6), at 4836.2 and 4865.2~Hz, are defect states bound to the removed corner. c) The band spectrum of the periodic $5\times5$ lattice (Floquet unit cell evaluated at $k = 2\pi n/5$) is cleanly gapped (4792--4906~Hz), with no modes inside the gap. d) The acoustic pressure profiles of the in-gap modes in a). e) Acoustic pressure profiles of the in-gap modes in b). All three-dimensional views share one orientation: the origin $(0,0)$ is the back column, the $x_1$ and $x_2$ directions point towards the viewer, and the corner columns are labeled.}
	\label{Fig:CornerModes}
\end{figure*}

Fig.~\ref{Fig:Hybridization} reports the hybridization diagrams as functions of the width of the bridges, for the three chosen lengths. As one can see, the frequency splittings are symmetric relative to $f_0$, assuring us that these bridges implement pure hoppings. Furthermore, $\Delta f$ increases from $0$ monotonically with the width of the acoustic bridge. This enables us to identify the widths $W_i$ of the bridges that implement coupling strengths 1 and 1/2, in some fixed frequency units. In the lattice, each resonator is shared by several bridges, which reduces every coupling to about 0.8 of its dimer value. We therefore re-tuned the widths once on the full lattice (final widths 2.17--2.67~mm) and trimmed each bridge length within its window. The coupling matrix reconstructed from the simulated lattice modes then matches Eq.~\eqref{Eq:TightBindingH}, including all coupling signs, with every coupling type within 2\% of its target and on-site offsets of 2.7~Hz rms, compared with a coupling strength of 40~Hz.

\section{Numerical Validation}

The finalized COMSOL model, a $5\times 5$ lattice of 100 H-resonators, can be seen in Fig.~\ref{Fig:CornerModes}, together with its simulated resonant spectrum under various conditions. Its repeating unit cell, shown in Fig.~\ref{Fig:UnitCell}, consists of one column of four stacked H-resonators, the two on-site bridges between them, and the eight bridges that connect it to the neighboring cells in the $x_1$ and $x_2$ directions. This should be compared with schematic of the couplings in Fig.~\ref{Fig:Couplings}, derived from the tight-binding Hamiltonian.

First, as predicted, a simulation with periodic boundary conditions returns a clean gapped spectrum, assuring us the bulk phase has the expected characteristic (Fig.~\ref{Fig:CornerModes}(c)); this calculation applies Floquet-periodic boundary conditions to the unit cell of Fig.~\ref{Fig:UnitCell}. Secondly, in an open configuration, the model exhibits mid-gap modes localized at the corners, with resonant frequencies within 2.5~Hz of $f_0$, while the bulk gap extends $\pm 43$~Hz around it (Figs.~\ref{Fig:CornerModes}(a,d)). Each of the four corners of the sample hosts one mode: the corners on the mirror diagonal host modes on orbitals 1 and 2, and the two off-diagonal corners host modes on the combination of orbitals 3 and 4, as predicted by the tight-binding model. Indeed, up to hybridizations which are ultrasensitive to small details and are impossible to control, the zero modes seen in the COMSOL simulation reported in Figs.~\ref{Fig:CornerModes}(a,d) are very similar to the ones reported in Fig.~\ref{Fig:TBP}.

Lastly, we demonstrate that the corner can be modified yet, as long as we preserve the mirror symmetry, the mid-gap corner modes endure (Figs.~\ref{Fig:CornerModes}(b,e)). Removing the corner cell $(0,0)$ moves the $(0,0)$ corner mode onto orbital 1 of the neighbouring cells $(0,1)$ and $(1,0)$, where it stays mid-gap (4851.5~Hz), exactly as the tight-binding model predicts, while the other three corner modes shift by less than 1~Hz. The perturbation also lifts the near-degeneracy of the $(0,0)$ and $(4,0)$ modes seen in Fig.~\ref{Fig:CornerModes}(a), so that each corner mode in Fig.~\ref{Fig:CornerModes}(b) sits on a single corner. Two further states, at 4836 and 4865~Hz, are bound to the defect itself and remain well separated from the corner modes.

 \end{document}